\documentclass{ieeetj}
\usepackage{fix-cm}
\newcommand{\seclogo}{}
\usepackage{cite}
\usepackage{amsmath,amssymb,amsfonts,amsthm}
\usepackage{mathtools}
\usepackage{graphicx,color}
\usepackage{textcomp}
\usepackage{xcolor}
\usepackage{bm}
\usepackage{mathdots}

\usepackage[caption=false, font=footnotesize]{subfig}
\usepackage{booktabs}
\usepackage{hyperref}
\usepackage{algorithm}
\usepackage{algpseudocode}
\usepackage{tikz}
\usetikzlibrary{arrows.meta,angles,shapes.geometric,quotes} 
\usepackage[normalem]{ulem}
\usepackage{cancel}

\usepackage{multirow}
\usepackage[dvipsnames]{xcolor} 
\usepackage{hyperref}
\hypersetup{colorlinks=true,linkcolor=RubineRed,citecolor=ForestGreen}

\let\labelindent\relax
\usepackage{enumitem}
\usepackage{tcolorbox}
\renewcommand{\thesubsection}{\thesection-\Alph{subsection}}

\newtheorem{lemma}{Lemma}

\newcommand{\nrf}{N_{\text{RF}}}

\newcommand{\km}{K_{\text{M}}}
\newcommand{\ceil}[1]{\left \lceil #1 \right \rceil}
\newcommand{\floor}[1]{\left \lfloor #1 \right \rfloor}

\newcommand{\Sm}{\bm S_m}

\newcommand{\scl}{\bm{s}_{\text{cl}}}

\newcommand{\Rx}{\bm R_{x}}
\newcommand{\rx}{\bm r_{x}}

\newcommand{\real}[1]{\text{Re}\left[#1\right]}
\newcommand{\imag}[1]{\text{Im}\left[#1\right]}

\newcommand{\nx}{N_{x}}
\newcommand{\ny}{N_{y}}
\newcommand{\nrfx}{N_{\text{RF},x}}
\newcommand{\nrfy}{N_{\text{RF},y}}
\newcommand{\Km}{K_{M}}

\newcommand{\Pfm}{\bm{P}_{f,m}}
\newcommand{\Sx}{\bm{S}_x}

\newcommand{\hatscl}{\hat{\bm{s}}_{\text{cl}}}
\newcommand{\hatrx}{\hat{\bm{r}}_x}

\newcommand{\Reps}{\bm{R}_{\epsilon}}

\newcommand{\hatReps}{\hat{\bm{R}}_{\epsilon}}
\newcommand{\hatRepsm}{\hat{\bm{R}}_{\epsilon,m}}
\newcommand{\hatSm}{\hat{\bm{S}}_m}
\newcommand{\Sigmacl}{\bm{\Sigma}_{\text{cl}}}

\newcommand{\we}{\bm{w}_e}
\newcommand{\wo}{\bm{w}_o}

\newcommand{\Sxlm}{\bm{S}_{x,\ell,m}}
\newcommand{\Sylm}{\bm{S}_{y,\ell,m}}

\newcommand{\Bm}{\bm{B}_m}

\newcommand{\rxl}{\bm{r}_{x,\ell}}
\newcommand{\ryl}{\bm{r}_{y,\ell}}

\newcommand{\rxy}{\bm{r}_{xy}}
\newcommand{\hatrxy}{\hat{\bm{r}}_{xy}}

\renewcommand{\vec}{\text{vec}} 

\newcommand{\Ixm}{\bm{\mathcal{I}}_{x,m}}
\newcommand{\Iym}{\bm{\mathcal{I}}_{y,m}}
\renewcommand{\Im}{\bm{\mathcal{I}}_m}

\newcommand{\Fx}{\bm{F}_x}
\newcommand{\Fy}{\bm{F}_y}

\newcommand{\Pf}{\bm{P}_f}

\newcommand{\tr}[1]{\text{tr}[#1]}

\newcommand{\alphal}{\bm{\alpha}_l}

\newcommand{\hatp}{\hat{\bm p}}

\newcommand{\blkdiag}{\text{blkdiag}}

\newcommand{\psil}{\psi_{\ell}}

\newcommand{\ax}{\bm{a}_x}
\newcommand{\Ax}{\bm{A}_x}

\newcommand{\psixl}{\psi_{x,\ell}}
\newcommand{\psiyl}{\psi_{y,\ell}}
\newcommand{\Rxl}{\bm R_{x,\ell}}
\newcommand{\Ryl}{\bm R_{y,\ell}}

\newcommand{\Deltarf}{\Delta_{\text{RF}}}

\newcommand{\alphauv}{\alpha_{u,v}}

\newcommand{\rdft}{r_{\text{dft}}}
\newcommand{\rwa}{r_{\text{wa}}}
\newcommand{\rank}{\text{rank}}
\newcommand{\mx}{M_x}
\newcommand{\my}{M_y}

\newcommand{\Omegax}{\bm{\Omega}_x}
\newcommand{\Omegay}{\bm{\Omega}_y}
\newcommand{\Sigmaclx}{\bm{\Sigma}_{\text{cl},x}}
\newcommand{\Sigmacly}{\bm{\Sigma}_{\text{cl},y}}
\newcommand{\Phix}{\bm{\Phi}_{x}}
\newcommand{\Phiy}{\bm{\Phi}_{y}}
\newcommand{\Pfmx}{\bm{P}_{f,m,x}}
\newcommand{\Pfmy}{\bm{P}_{f,m,y}}
\newcommand{\Pfxy}{\bm{P}_{f,xy}}
\newcommand{\Pfx}{\bm{P}_{f,x}}
\newcommand{\Pfy}{\bm{P}_{f,y}}
\newcommand{\Psixy}{\bm{\Psi}_{xy}}
\newcommand{\Omegaxy}{\bm{\Omega}_{xy}}
\newcommand{\Sigmaclxy}{\bm{\Sigma}_{\text{cl},xy}}
\newcommand{\Phixy}{\bm{\Phi}_{xy}}
\newcommand{\Pfmxy}{\bm{P}_{f,m,xy}}

\def\BibTeX{{\rm B\kern-.05em{\sc i\kern-.025em b}\kern-.08em
    T\kern-.1667em\lower.7ex\hbox{E}\kern-.125emX}}

\makeatletter
\newcommand\fs@ruled@black{%
  \def\@fs@cfont{\bfseries}\let\@fs@capt\floatc@ruled
  \def\@fs@pre{\color{black}\hrule height.8pt depth0pt \kern2pt}%
  \def\@fs@post{\color{black}\kern2pt\hrule\relax}%
  \def\@fs@mid{\color{black}\kern2pt\hrule\kern2pt}%
  \let\@fs@iftopcapt\iftrue}
\makeatother

\floatstyle{ruled@black}
\restylefloat{algorithm}

\def\BibTeX{{\rm B\kern-.05em{\sc i\kern-.025em b}\kern-.08em
    T\kern-.1667em\lower.7ex\hbox{E}\kern-.125emX}}
\AtBeginDocument{\definecolor{tmlcncolor}{cmyk}{0.93,0.59,0.15,0.02}\definecolor{NavyBlue}{RGB}{0,86,125}}

\def\OJlogo{\vspace{-4pt}$<$Society logo(s) and publication title will appear here.$>$}
\def\seclogo{\vspace{10pt}$<$Society logo(s) and publication title will appear here.$>$}

\def\authorrefmark#1{\ensuremath{^{\textbf{#1}}}}

\begin{document}
\receiveddate{XX Month, XXXX}
\reviseddate{XX Month, XXXX}
\accepteddate{XX Month, XXXX}
\publisheddate{XX Month, XXXX}
\currentdate{XX Month, XXXX}
\doiinfo{XXXX.2022.1234567}

\title{Efficient DoA Estimation for Linear and Rectangular Arrays with Hybrid Architectures Using Compact DFT Codebooks
\\
}

\author{Miguel Rivas-Costa\authorrefmark{1}, Carlos Mosquera\authorrefmark{1}}
\affil{atlanTTic Research Center, Universidade de Vigo, Vigo, Spain}
\corresp{Corresponding author: Miguel Rivas-Costa (email: mrivascosta@gts.uvigo.es).}
\authornote{This research was funded in part by MICIU/AEI/10.13039/501100011033 FEDER/EU under projects PID2021-122483OA-I00 and PID2022-136512OB-C22. Work supported in part by the Xunta de Galicia (Secretaría Xeral de Universidades) under a Predoctoral Scholarship (cofunded by the European Social Fund) ED481A 2022/407.}

    

\begin{abstract}
Hybrid Analog and Digital (HAD) architectures significantly reduce hardware overhead but introduce severe dimensionality compression, which strips the Spatial Covariance Matrix (SCM) of the degrees  of freedom required for high-resolution Direction-of-Arrival (DoA) estimation. This challenge is further compounded by passive Butler-matrix implementations of Discrete Fourier Transform (DFT) analog beamforming, which avoid active phase shifters and amplifiers.
 In this paper, we propose a Generalized Least Squares (GLS) framework that exploits the Cauchy-like displacement structure that arises after DFT beamforming. By leveraging this structure, we develop a highly efficient numerical technique to recover the SCM for uniform linear arrays with a complexity of $\mathcal{O}(\nrf^2\nx)$, where $\nx$ is the number of antennas and $\nrf$ the number of RF-chains. Simulations demonstrate that our estimator approaches the Cramér-Rao Bound (CRB) while outperforming state-of-the-art methods.
\end{abstract}


\maketitle

\section{INTRODUCTION}
\subsection{MOTIVATION}
The ability to resolve the spatial origin of impinging wavefronts is a critical requirement for various applications, including underwater sensing \cite{wang2024}, advanced radar architectures \cite{sorkhabi2025}, or autonomous vehicle implementations \cite{su2025}. Formally addressed as  Direction-of-Arrival (DoA) estimation, this problem centers on characterizing the spatial components of the incident wave to resolve the angular coordinates of multiple sources \cite{krim1996}. The DoA estimation problem is well established in the array signal processing literature, where powerful algorithms have been developed that achieve high performance with moderate computational complexity \cite{trees2002}. Nevertheless, the vast majority of existing studies focus on fully-digital (FD) architectures, in which each antenna is connected to a dedicated radio-frequency (RF) chain \cite{liu2023,krim1996,pesavento2023,trees2002,tuncer2009classical}. Assigning one RF chain per antenna is infeasible due to physical constraints, cost, and power consumption \cite{doan2004}. To address this challenge, Hybrid Analog and Digital (HAD) architectures offer a cost-effective solution, in which antenna outputs are first processed by an analog combiner, reducing dimensionality, and then passed to a smaller set of RF chains for digitization \cite{heath2016,sun2014,gao2018,venkateswaran2010}.

Depending on the design of the analog combiner, three HAD main categories can be identified: Fully-Connected (FC), Partially-Connected (PC), and Switch-Based (SB) architectures \cite{heath2016}. FC and PC arrays employ networks of phase shifters in the analog stage, enabling high beamforming gains but at the expense of greater cost and power consumption. In contrast, SB architectures achieve lower cost and energy requirements by digitizing only a subset of the antenna elements, though this comes at the price of reduced beamforming flexibility \cite{mendezrial2016}. To combine the efficiency of switch-based solutions with the beamforming capabilities of phase-shifter designs, recent efforts have considered architectures that integrate highly efficient hardware structures. Butler Matrices  provide an attractive hardware implementation because they realize the spatial Discrete Fourier Transform (DFT) in a passive and highly efficient manner \cite{butler1961,vallappil2021}. They can be interpreted as a specific form of passive FC-HAD architecture, where the phase shifts are fixed, restricting the synthesized beams to a predefined set of angular directions. While this limitation reduces beamforming flexibility compared to conventional phase-shifter networks, Butler Matrices offer significant advantages in terms of cost and robustness against hardware impairments, such as insertion losses \cite{garciarodriguez2016}. These practical benefits have motivated the development of novel beamforming techniques specifically tailored to exploit Butler Matrix based architectures, rather than conventional HAD structures. In \cite{noh2017,han2018}, the authors developed algorithms that leverage DFT beams to enhance the data rate in scenarios dominated by a single propagation path. Similarly, \cite{wu2019expeditius} proposed a method for estimating the one dominant path in presence of multipath, whereas the authors in \cite{haardt2024} employed DFT beams to jointly estimate DoAs and phase errors within a priori constrained angular regions using Uniform Linear Arrays (ULAs), and this approach was subsequently extended to Uniform Rectangular Arrays (URAs) in \cite{haardt2025}. Finally, \cite{wu2023green} discussed the challenges and opportunities associated with DFT-based architectures for joint sensing and communication.
\subsection{BACKGROUND}
In the literature, most contributions addressing estimation in HAD arrays are framed in the context of channel estimation, where DoA typically appears as a subtask within the overall procedure. Channel estimation methods generally rely on the transmission of known pilot sequences, which simplifies the estimation problem by providing structured a priori information \cite{lin2021, berberidis2020, chuang2015, chiu2019, alkhateeb2014, noh2022}. In contrast, the DoA estimation literature often adopts a more general and challenging scenario in which no pilots are assumed, and the received signals are treated as completely unknown \cite{liu2023,pesavento2023}. This pilot-free setting requires dedicated estimation techniques and has motivated a distinct body of research in array signal processing. Moreover, DoA estimation with hybrid architectures must explicitly account for both the geometry of the antenna array and the structure of the analog combiner. Each configuration, whether FC, PC, or SB, has its own particularities, and techniques developed for one setup are not directly transferable to another. As a result, the proposals  in the  literature have diversified into several algorithmic families, each exploiting different structural properties of the system:

\begin{itemize}[leftmargin=*]
    \item [i)] Compressed sensing (CS) methods:  \cite{wang2009,gu2011,zhou2017,ganguly2022,kilic2022,ibrahim2017,wang2025}. CS-based techniques are highly valued for estimation in low-snapshot regimes. However, these techniques face significant problems. They require solving complex sparse recovery optimization problems, leading to computationally intensive steps. 
    Furthermore, standard CS formulations require a discretization of the angular search space; this introduces grid-mismatch errors, that potentially lead to severe estimation errors when the true DoA does not align perfectly with the predefined grid points.

    
    \item [ii)] Phase-Alignment methods: \cite{huang2010,li2019,li2021,zhang2015,zhang2016,wu2018,shu2018,zhan2024,wang2023,shu2025}. Characterized by their appealingly low computational complexity, they are fundamentally constrained to environments with only a single impinging source, which is often an impractical assumption in realistic applications.
    

    \item[iii)] Alternative approaches have also been proposed in the literature. For example, the authors in \cite{matiwax1999} introduced a method based on alternating projections that can be applied to any uniform linear array architecture and demonstrated that it is asymptotically efficient. However, they did not address the design of the beamforming matrices or the codebook structure, leaving this aspect as an open problem. More recently, \cite{zhang2022} developed a Maximum Likelihood Estimator (MLE) for PC, FC, and SW architectures, where the approach requires pseudorandom beamforming. In \cite{abdelbadie2024}, an iterative method was proposed in which estimation is progressively refined using DFT codebooks. Additional recent contributions include \cite{liu2025}, which addresses PC arrays with non-uniform connections, and \cite{hoang2025}, which investigates architectures based on lens antennas.
\end{itemize}

Special attention must be  paid to the different performance bounds that have been derived to benchmark the impact of DoA estimation algorithms. In \cite{sheinvald1997}, the authors provided the Cramér–Rao Bound (CRB) for the unconditional model in SB architectures, where the transmitted signal is modeled as a complex Gaussian random process, and analyzed the estimator behavior when the number of RF chains is smaller than the number of sources. The CRB for FC and PC architectures under the unconditional model was presented in \cite{matiwax1999}. 
In \cite{rivas2023direction}, the corresponding Ziv–Zakai Bound (ZZB) was established, offering a tighter benchmark in low-SNR regimes. 
More recently, \cite{zhang2022} derived the CRB for the conditional model, where the transmitted signal is treated as an unknown deterministic vector to be estimated, for PC, FC, and SB architectures.

All of the above approaches are usually subject to limiting assumptions, such as the \textit{a priori} knowledge of the number of  impinging signals, constrained to a single source in many cases. While this assumption is reasonable in FD arrays due to the extensive body of work on source enumeration \cite{matiwax1989}, no equivalent methods exist for hybrid architectures. To circumvent this limitation, a new research direction reformulated the DoA problem as a second-order estimation task, in which the spatial covariance matrix (SCM) of an equivalent FD array is reconstructed from the low-dimensional measurements collected by the hybrid array. Once the covariance is estimated, classical FD techniques can be directly applied. This idea was first explored in \cite{ariananda2012} for SB architectures with correlated signals, where a variant of MUSIC was employed for estimation. The approach was later extended to phase-shifter structures in \cite{li2020} through the Beam Sweeping Algorithm (BSA) for single-RF architectures, and in \cite{zhou2024}, where one-bit measurements were used. BSA was further generalized to multi-RF PC-HAD arrays in \cite{yan2021,liu2021} and to FC-HAD architectures in \cite{fu2020}. Subsequent efforts sought to reduce the computational burden of covariance reconstruction. In \cite{yan2021_nrf1}, an equivalent real-valued covariance matrix was estimated to reduce computations, while \cite{zhou2024_subarray} introduced a subarray-based implementation in which smaller covariance matrices are estimated, significantly lowering complexity. Most of these works assume DFT-based beamforming, yet the required codebook size has rarely been analyzed. This question is critical in highly dynamic environments, where fast covariance recovery is essential. In this context, \cite{shmonin2023} showed that for single-RF architectures, a codebook of size twice the number of antennas suffices for exact covariance reconstruction, with the Fast Fourier Transform (FFT) providing a drastic reduction in computational cost. Building on this line of work, our previous contribution \cite{rivas2025} demonstrated that by jointly exploiting both the measured power and the correlations between RF-chain outputs, the codebook size can be reduced by up to 50\% for certain array geometries. Despite these advances, existing approaches achieve only limited performance 
because they do not account for the statistics of the estimation errors when reconstructing the SCM. This limitation motivated preliminary investigations for two-dimensional arrays in \cite{covrec_2D_rivas2025}. Subsequently,  \cite{rivas2026} introduced a Generalized Least Squares (GLS) estimator specifically tailored to one-dimensional Partially-Connected (PC) architectures. Extending this framework to FC architectures is considerably more challenging, as the underlying hardware mapping fundamentally changes the estimation problem and requires the development of new codebook designs for SCM reconstruction. Nevertheless, as will be shown, the resulting FC-induced statistics exhibit a displacement structure that enables a computationally efficient solution.


\subsection{CONTRIBUTIONS}
In this paper, we propose a high-resolution DoA estimation algorithm based on a Generalized Least Squares (GLS) framework for Spatial Covariance Matrix (SCM) estimation. We specifically focus on hardware-efficient Fully-Connected (FC) architectures, where the analog beamforming is performed by a passive Butler Matrix  via the DFT, with the resulting signal compressed through a switching network. 

To guarantee the accurate recovery of the SCM, we develop a specialized codebook that ensures reconstruction by exploiting the Cauchy-like displacement structure \cite{kailath1995displacement}  that arises  after DFT-based beamforming, applicable to both Uniform Linear Arrays (ULAs) and Uniform Rectangular Arrays (URAs). 

By leveraging this Cauchy-like displacement structure, we design a computationally efficient algorithm to estimate the SCM. Specifically, for an array with $\nx$ antennas and $\nrf$ RF chains, the proposed approach achieves a remarkably low computational complexity of $\mathcal{O}(\nrf^2\nx)$. This highly efficient matrix estimation facilitates integration of subsequent high-resolution, subspace-based DoA estimation techniques, which typically incur a baseline complexity of $\mathcal{O}(\nx^3)$. As numerical simulations demonstrate, this framework offers a substantial computational advantage over alternative methods that require exhaustive grid searches to achieve comparable estimation performance.

\subsection{NOTATION}
Boldface uppercase (lowercase) letters denote matrices (vectors). The operators $(\cdot)^{T}$ and $(\cdot)^{H}$ represent transpose and Hermitian (conjugate transpose), respectively, and $\bm I_{N}$ denotes the $N \times N$ identity matrix. The Frobenius and $\ell_{2}$ norms are denoted by $||\cdot||_{F}$ and $||\cdot||_{2}$, respectively. The real and imaginary parts of a complex quantity are written as $\real{\cdot}$ and $\imag{\cdot}$. The ceiling operator $\lceil \cdot \rceil$ denotes the smallest integer greater than or equal to its argument. For a matrix $\bm A$ of size $m\times n$, the entry in the $u$th row and $v$th column is written as $\bm A[u,v]$, where the index range over $0\leq u< m$ and $0\leq v < n$. The vector $\bm e_{u}^{N}$ is the $N\times 1$ vector, with all entries equal to zero except for a one in the  $u$th position. The Kronecker product of two matrices $\bm A$ and $\bm B$ is denoted $\bm A \otimes \bm B$. Finally, $(a)_N$ denotes the remainder of the division of $a$ by $N$.

\section{ULA: SIGNAL MODEL}\label{sec:alg_ulas}
We consider $L$ uncorrelated narrowband radio signals with wavelength $\lambda$ that impinge on an antenna  array\footnote{Broadband signals or signals at different carrier frequencies can be also accommodated by operating with discrete frequency bins \cite{trees2002}.}, equipped with $\nx$ antennas with inter-element spacing $d_x=\lambda/2$. As illustrated in Fig.~\ref{fig:ula_geometry}, the antennas are connected to a passive hardware-efficient analog beamforming network. In this work, this network is a Butler Matrix, which performs a spatial transformation of the incoming signal $\bm x(t)\in\mathbb{C}^{\nx\times 1}$ via the DFT $\Fx^H\bm x(t)$, with $\Fx$ defined as 
\begin{align}\label{eq:DFT}
    \Fx[u,v] \triangleq\frac{1}{\sqrt{\nx}}e^{j\frac{2\pi}{\nx}uv}\quad 0\leq u,v<\nx.
\end{align}
\noindent To reduce the number of Radio Frequency (RF) chains, a switch network follows the passive device to select a subset of $\nrf\leq \nx$ outputs before down-conversion and sampling. We assume that the system employs $M$ different switch configurations, represented by a set of $M$ selection matrices $\{\Ixm\in\{0,1\}^{\nx\times \nrf}\}$. The effective beamforming matrix for the $m$-th configuration is defined by $\Bm = \Fx\Ixm$, and  encapsulates the composite operation of the Butler Matrix and the switch network. While the collection of the beamforming matrices $\{\Bm\}_{m=1}^{M}$ conventionally constitutes the system \textit{codebook}, we shall also use the term to refer to the underlying set of switch configuration $\{\Ixm\}_{m=0}^{M-1}$. Each configuration is maintained constant during a time window consisting of $\Km$ time snapshots, hereafter denoted as \textit{batch}. Let $\bm x(t)\triangleq \bm A_x(\bm \psi)\bm s(t)+\bm n(t)$ be the received signal in the FD array. In the hybrid architecture, after down-conversion and sampling,  the $m$th batch signal reads 
\begin{align}\label{eq:ymt}
    \bm y_m(t) = \bm B_m^H\bm x(t+m\Km)
\end{align}
where $0\leq t<\Km$ and $0\leq m<M$. The transmitted signal vector $\bm s(t)\in\mathbb{C}^{L\times 1}$ is modeled as a zero-mean random process with diagonal covariance matrix $\bm R_s = \text{diag}(\sigma_{s,1}^2,\dots,\sigma_{s,L}^2)$. The additive noise $\bm n(t)\in\mathbb{C}^{\nx\times 1}$ is modeled as zero-mean complex Gaussian  with covariance matrix $\sigma_n^2\bm I_{N}$. The array manifold matrix $\Ax(\bm \psi) \triangleq [\ax(\psi_1),\dots,\ax( \psi_{L})] \in \mathbb{C}^{\nx\times L}$ aggregates the steering vectors of the $L$ sources. Under ideal calibration, and defining the spatial frequency as $\psixl\triangleq \pi \sin\theta_\ell$, the steering vector for the $\ell$-th source is given by
\begin{equation}
\ax(\psil) \triangleq \begin{bmatrix}
1 & \dots & e^{j(\nx-1)\psixl}
\end{bmatrix}^T.
\end{equation}
After beamforming, the received signal vector in the $m$-th batch, $\bm y_m(t)\in\mathbb{C}^{\nrf\times 1}$, is a zero-mean random process with covariance matrix $\Sm \triangleq \mathbb{E}[\bm y_m(t)\bm y_m(t)^H]$ given by
\begin{equation}\label{eq:Sm}
    \Sm =\Bm^H\Rx\Bm = \Ixm^T\Sx\Ixm,
\end{equation}
where  $\Sx\triangleq \Fx^H\Rx\Fx$. The matrix $\Rx\triangleq \mathbb{E}[\bm x(t)\bm x(t)^H]=\bm A_x(\bm \psi)\bm R_s\bm A(\bm \psi)^H+\sigma_n^2\bm I_{\nx}$, is the Spatial Covariance Matrix (SCM) of the equivalent FD. From  \eqref{eq:Sm}, we can readily see that  $\Sm$ consists of samples of the spectral representation $\Sx$ of the SCM. Moreover, since the source signals are uncorrelated, $\Rx$ exhibits a Hermitian Toeplitz structure and  can be fully characterized by $2\nx-1$ unique real values. We define the \textit{spatial covariance sequence} $\rx\in\mathbb{C}^{(2\nx-1)\times 1}$ such that its $q$-th entry relates to the elements of the SCM according to
\begin{equation}
\label{eq:rx}
    \rx[q] \triangleq \Rx[q,0], \; 0\leq q \leq \nx-1
\end{equation}
with $\rx[q] \triangleq \rx^*[-q]$ for $-\nx+1 \leq q <0$.

\begin{figure}[t!]
    \centering
    \begin{tikzpicture}[>=Stealth, scale=0.7, transform shape]
        
        \coordinate (S0) at (-3, 7.5);
        \coordinate (SL) at (2, 7.5);
        \coordinate (Focus) at (0, 6);        
        \draw[thick, dashed] (Focus) -- (0, 8);
        
        \fill (S0) circle (2.5pt) node[above=5pt] {$s_0(t + mK_M)$};
        \fill (SL) circle (2.5pt) node[above=5pt] {$s_{L-1}(t + mK_M)$};
        
        \draw[->, thick] (S0) -- (Focus);
        \draw[->, thick] (SL) -- (Focus);
        
        \draw[->] (0, 7) arc (90:125:2) node[midway, above left] {$\theta_0$};
        \draw[->] (0, 7) arc (90:62:2) node[midway, above right] {$\theta_{L-1}$};

        \newcommand{\beam}[2]{
            \begin{scope}[rotate=#1]
                \filldraw[fill=#2, draw=black, thick] 
                    (0,0) .. controls (0.5, 0.3) and (2.5, 0.6) .. (3, 0) 
                          .. controls (2.5, -0.6) and (0.5, -0.3) .. (0,0);
            \end{scope}
        }
        
        \begin{scope}[yshift=3cm]
            \beam{160}{gray!20} \beam{140}{gray!20} \beam{120}{gray!20}
            \beam{100}{gray!20} \beam{80}{gray!20}              
            \beam{20}{gray!20}
            
            \draw[thick, black, rotate around={50:(0,0)}] (2.2, 1.2) arc (90:270:0.4 and 1.2);
            
            \beam{60}{green!60!black!50} 
            \beam{40}{red!80}            
            
            \draw[thick, black, rotate around={50:(0,0)}] (2.2, -1.2) arc (-90:90:0.4 and 1.2);
            

            \node[rotate=-40] at (50:2.9) {$\cdots$};
            \node at (50:3.4) {$N_{\text{RF}}$};
        \end{scope}
        
        \def\antY{1.8} 
        \newcommand{\antenna}[1]{
            \begin{scope}[shift={(#1,\antY)}]
                \draw[thick] (0,0) -- (0,0.05); 
                \draw[thick, fill=yellow!30] (0,0.05) -- (-0.2,0.3) -- (0.2,0.3) -- cycle;
            \end{scope}
        }
        \antenna{-2.2} \antenna{-1.4} \antenna{1.4} \antenna{2.2} 
        
        \node at (-1.8, \antY+0.15) {$\cdots$};
        \node at (0, \antY+0.15) {$\cdots$};
        \node at (1.8, \antY+0.15) {$\cdots$};
        
        \node[left] at (-2.3, \antY+0.15) {$1$};
        \node[right] at (2.3, \antY+0.15) {$N_x$};
        \node[above] at (0, \antY+0.35) {$\bm x(t + mK_M)$};

        \draw[thick] (-2.2, \antY) -- (-2.2, \antY-0.2);
        \draw[thick] (-1.4, \antY) -- (-1.4, \antY-0.2);
        \draw[thick] (1.4, \antY) -- (1.4, \antY-0.2);
        \draw[thick] (2.2, \antY) -- (2.2, \antY-0.2);

        \draw[thick, fill=blue!15] (-2.6, \antY-0.2) rectangle (2.6, \antY-1.5);
        \node at (0, \antY-0.85) {\scalebox{1.2}{$\mathbf{F}_x^H$}};
        \node[right, align=left] at (2.7, \antY-0.85) {BUTLER\\MATRIX};

        \def\midY{\antY-1.4}

        \draw[thick, fill=blue!15] (-2.6, \midY-1.2) rectangle (2.6, \midY-2.5);
        \node at (0, \midY-1.85) {\scalebox{1.2}{$\boldsymbol{\mathcal{I}}_{x,m}$}};
        \node[right, align=left] at (2.7, \midY-1.85) {SWITCH\\NETWORK};
        
        \draw[thick] (-2.2, \antY-1.5) -- (-2.2, \midY-1.2);
        \draw[thick, green!60!black] (-1.4, \antY-1.5) -- (-1.4, \midY-1.2); 
        \draw[thick, red!80] (1.4, \antY-1.5) -- (1.4, \midY-1.2);           
        \draw[thick] (2.2, \antY-1.5) -- (2.2, \midY-1.2);

        \draw[fill=white, thick] (-2.2, \antY-1.5) circle (2pt);
        \draw[fill=white, thick, draw=green!60!black] (-1.4, \antY-1.5) circle (2pt);
        \draw[fill=white, thick, draw=red!80] (1.4, \antY-1.5) circle (2pt);
        \draw[fill=white, thick] (2.2, \antY-1.5) circle (2pt);
        
        \node at (-1.8, \antY-1.8) {$\cdots$};
        \node at (0, \antY-1.8) {$\cdots$};
        \node at (1.8, \antY-1.8) {$\cdots$};
        
        \node at (0, \midY-0.7) {$\bm F_x^H \bm x (t + mK_M)$};

        \draw[fill=white, thick] (-2.2, \midY-1.2) circle (2pt);
        \draw[fill=white, thick, draw=green!60!black] (-1.4, \midY-1.2) circle (2pt);
        \draw[fill=white, thick, draw=red!80] (1.4, \midY-1.2) circle (2pt);
        \draw[fill=white, thick] (2.2, \midY-1.2) circle (2pt);
        
        \node at (-1.8, \midY-1) {$\cdots$};
        \node at (0, \midY-1) {$\cdots$};
        \node at (1.8, \midY-1) {$\cdots$};
        
        \draw[<-, thick, green!60!black] (-1.4, \midY-1.2) -- (-1.5, \midY-2.2);
        \fill[green!60!black] (-1.5, \midY-2.2) circle (2pt);
        
        \draw[<-, thick, red!80] (1.4, \midY-1.2) -- (1.5, \midY-2.2);
        \fill[red!80] (1.5, \midY-2.2) circle (2pt);

        \def\rfY{\midY-3.2}

        \draw[thick, fill=purple!20] (-2.3, \rfY+0.45) rectangle (-0.7, \rfY-0.45) node[midway] {RF-chain};
        \draw[thick, fill=purple!20] (0.7, \rfY+0.45) rectangle (2.3, \rfY-0.45) node[midway] {RF-chain};

        \draw[thick, green!60!black] (-1.5, \midY-2.2) -- (-1.5, \rfY+0.45);
        \draw[thick, red!80] (1.5, \midY-2.2) -- (1.5, \rfY+0.45);
        
        \node at (0, \rfY) {$\cdots$};
        \node[below] at (0, \rfY-0.2) {$N_{\text{RF}}$};

        \draw[thick] (-1.5, \rfY-0.45) -- (-1.5, \rfY-0.8);
        \draw[thick] (1.5, \rfY-0.45) -- (1.5, \rfY-0.8);
        
        \node at (0, \rfY-1.2) {\scalebox{1.2}{$\bm{y}_m(t) = \bm{\mathcal{I}}_{x,m}^T \bm{F}_x^H \bm{x}(t + mK_M)$}};
    \end{tikzpicture}
    \caption{$L$ sources impinge upon an ULA of $\nx$ antennas and $\nrf$ RF-chains. The received signal is processed first by a Butler Matrix and then compressed by a network of switches. }
    \label{fig:ula_geometry}
\end{figure}


\section{ULA: CODEBOOK CHARACTERIZATION}\label{sec:codebook_design}
In this section we introduce the codebook of switch configurations used to guarantee the recovery of the spatial covariance sequence $\rx$ in \eqref{eq:rx}  from the set of low-dimensional batch covariance matrices $\{\Sm\}$. We begin in Section~\ref{sec:cauchylike_driven_codebook_design} by detailing the codebook that guarantee recovery by exploiting the
 Cauchy-like displacement structure that arises after DFT beamforming. Section~\ref{sec:codebook_design_num_measurements} analyzes the benefit of using all available complex samples for covariance recovery rather than traditional power-only solutions found in the literature.

\subsection{CAUCHY-LIKE DRIVEN CODEBOOK DESIGN}\label{sec:cauchylike_driven_codebook_design}
The design of a codebook is one of the most challenging aspects in HAD architectures. While significant literature exists regarding \textit{beamspace processing}, which assumes that signals lie within a known angular region \cite{trees2002}, we address  the more general scenario here, by means of judiciously chosen time-varying beamforming matrices in the absence of  \textit{a priori} knowledge of the origin of the signals, a problem that remains largely open. 

A basic requirement for any such codebook is that the SCM $\Rx$ can be recovered from the set of $M$ projections $\{\Sm=\Ixm^T\Sx\Ixm\}$, which provide access to specific sub-blocks of the 2D-DFT  matrix $\Sx = \Fx^H\Rx\Fx$.  If $\Fx[:,u]$ denotes the $u$-th column of $\Fx$, by exploiting the Hermitian Toeplitz structure of the SCM, where $\Rx[u,v]=\rx[v-u]$, the entry $\Sx[u,v]=\Fx[:,u]^H\Rx\Fx[:,v]$ can be expressed via the following spectral representation:
\begin{align}\label{eq:Sx_uv}
    &\Sx[u,v] =\frac{1}{\nx}\sum_{p=0}^{\nx-1}\sum_{q=0}^{\nx-1}\bm R_x[p,q]e^{-j\frac{2\pi}{\nx}(up-vq)}\nonumber\\
    &=\frac{1}{\nx}\sum_{q=-\nx+1}^{\nx-1}\bm{\mathsf w}(u,v,q)\rx[q] e^{-j\frac{2\pi}{\nx}(up-vq)}.
\end{align}
where the weight function is 
\begin{equation}
    \bm{\mathsf{w}}(u,v,q) \triangleq e^{-j\frac{2\pi}{\nx}uq}\sum_{p=\text{max}(0,q)}^{\text{min}(\nx-1,\nx-1-q)} e^{-j\frac{2\pi}{\nx}(u-v)(p-q)}.\nonumber
\end{equation}
From \eqref{eq:Sx_uv}, we observe that the entries of $\Sx$  constitute samples of the \textit{generalized spectrum} (or \textit{bifrequency spectrum) } of $\bm x(t)$  \cite{gerr1994,naghibi2012}, which characterizes the spatial covariance sequence $\rx$ in the spectral domain. Hence, designing a codebook involves selecting the probing frequencies to guarantee the recovery of $\rx$. 

As established in \cite{heining1998}, $\Sx$ exhibits a Cauchy-like displacement structure, with its entries satisfying
\begin{equation}\label{eq:Suv}
     \Sx[u,v] = \alpha_{u,v}\cdot\begin{cases}
         \text{Re}\left\{S_u-\tfrac{1}{\nx}S_v'\right\}\quad& u = v,\\
         \text{Im}\left\{S_u-S_v\right\}\quad &  u\neq v.
     \end{cases}
 \end{equation}

\begin{figure*}[ht] 
    \centering
    \subfloat[][\label{fig:codebook_beampattern}]{%
        \includegraphics[width=0.42\textwidth]{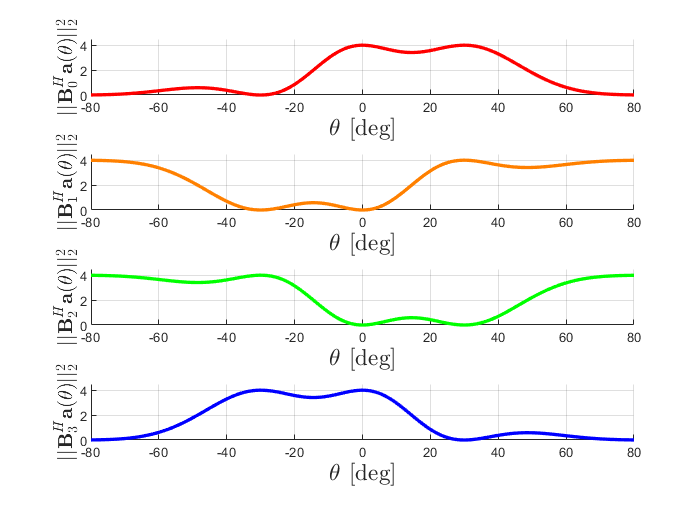}%
    }%
    \subfloat[][\label{fig:representation_S}]{%
        \includegraphics[width=0.42\textwidth]{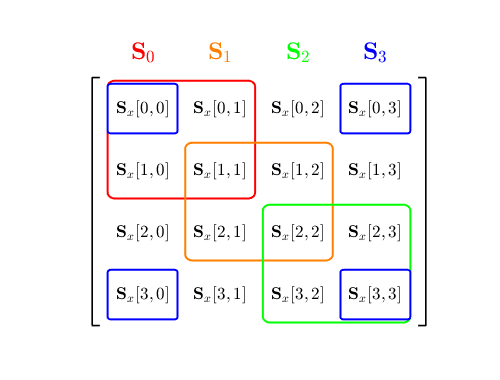}%
    }
    \vspace{-1em} 
    \subfloat[][\label{fig:codebook_crb}]{%
        \includegraphics[width=0.42\textwidth]{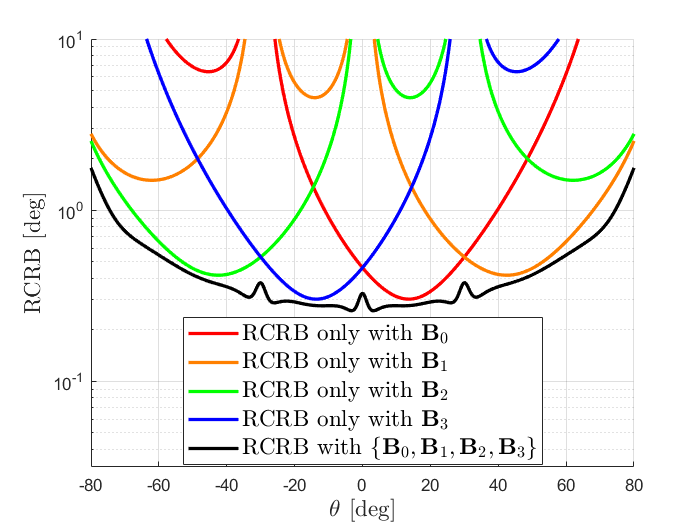}%
    }
    \subfloat[][\label{fig:Pf}]{%
        \includegraphics[width=0.42\textwidth]{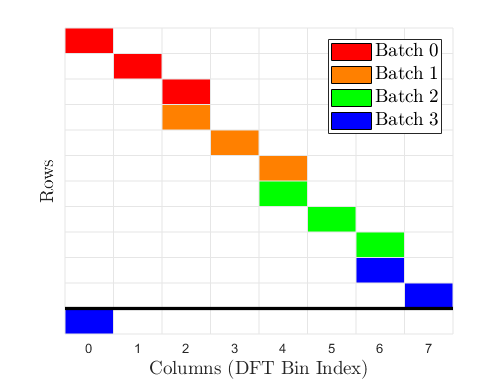}%
    }
    \caption{Array parameters: $\nx=4,\ \nrf=2,\ M=4$. The curve colors correspond to batch index $m\in\{0,1,2,3\}$, mapping respectively to red, orange, green and blue. (a) Synthesized beam patterns $\|\bm B_m^H\bm a(\theta)\|_2^2$ evaluated for each individual batch $m$. (b) Sampling of $\Sx$. (c) RCRB computed for each individual beamforming matrix $\{\Bm\}$ (colored curves) compared against the overall RCRB achieved by the entire codebook (black curve). (d) Visual representation of matrix $\Pf$, where the horizontal thick black line distinguishes the rows dedicated to sensing the $2\nx$-DFT frequencies from the row resulting from the spectral wrap-around.}
    \label{fig:full_analysis}
\end{figure*}

 \noindent The auxiliary spectral variables are defined as
 \begin{align}
     S_u &\triangleq \frac{\bm{r}_{x}[0]}{2} + \sum\nolimits_{q=1}^{N-1}\bm{r}_{x}[q]e^{-j\frac{2\pi}{\nx}uq},\\
     S'_u &\triangleq \sum\nolimits_{q=1}^{N-1}q\cdot \bm{r}_{x}[q]e^{-j\frac{2\pi}{\nx}uq},
 \end{align}
 and the scaling coefficients are given by
\begin{equation}\label{eq:alpha_uv}
    \alpha_{u,v} \triangleq  \begin{cases}
        \quad \quad \quad 2 & u=v,\\
        \frac{2j}{\nx}(1-e^{j\frac{2\pi}{\nx}(v-u)})^{-1}&u\neq v.
    \end{cases}
\end{equation} 
An important implication of the Cauchy-like displacement structure in \eqref{eq:Suv} is that the off-diagonal entries are not independent. Specifically, every off-diagonal element can be expressed as a linear combination of the first off-diagonal entries, with $|u-v|=1$, through the differences $\imag{S_u-S_v}$. As a consequence, the entire $\nx\times\nx$ matrix $\Sx$ is uniquely determined by a minimal set of $2\nx-1$ real-valued parameters,  $\nx$ and $\nx-1$  embedded in the main diagonal and the first off-diagonal, respectively. 

For convenience, we adopt  an augmented representation of $2\nx$ elements for subsequent derivations. To this end, we define the spectral vector
$\scl\in\mathbb{R}^{2\nx\times 1}$ entry-wise for $0\leq u<2\nx$ as
\begin{align}
    \scl[u] &\triangleq \begin{cases}
        \real{S_{u/2}-\frac{1}{\nx}S_{u/2}'}&u\; \text{even},\\
        \imag{S_{(u-1)/2} - S_{(u+1)/2}} &u\; \text{odd},
    \end{cases}\label{eq:scl}\\
    &=\sum_{q=-\nx+1}^{\nx-1}\begin{cases}
        \frac{1}{2}(1- \frac{|q|}{\nx})e^{-j\frac{\pi}{\nx}uq}\rx[q]&u\; \text{even},\\
        \sin(\frac{\pi|q|}{\nx})e^{-j\frac{\pi}{\nx}uq}\rx[q]&u\; \text{odd}.
    \end{cases}\nonumber
\end{align}
From a physical perspective, the even-indexed entries  $\{\scl[2u]\}_{u=0}^{\nx-1}$ represent the windowed power spectral density evaluated at $\nx$-DFT bins. In contrast, the odd-indexed entries $\{\scl[2u+1]\}_{u=0}^{\nx-1}$ capture the inter-frequency correlation between adjacent bins. 
This representation reveals a fundamental property of the DFT-based codebooks: although the hardware is limited to $\nx$-DFT beamforming, sampling the first off-diagonal of $\Sx$ effectively samples a $2\nx$-point DFT grid. Note that the final entry $\scl[2\nx-1]$ is a linear combination of the preceding odd-indexed elements, satisfying the constraint $\scl[2\nx-1]=-\sum_{u=0}^{\nx-2}\scl[2u+1]$. 

The vector $\scl$ is related to $\rx$ via the linear mapping:
\begin{equation}\label{eq:scl_mat}
    \scl = \bm \Phi\bm \Omega\rx,
\end{equation}
where $\bm \Omega\in\mathbb{C}^{(2\nx-1)\times (2\nx-1)}$ is the windowed $2\nx$-DFT matrix defined in Appendix~\ref{appendix:A2}.
The linear dependency of the final entry of $\scl$ is enforced by $\bm \Phi\in\{-1,0\}^{2\nx\times (2\nx-1)}$:
\begin{align}\label{eq:Phi}
    \bm \Phi &\triangleq [\bm I_{2\nx-1}, \bm v]^T,\quad
    \bm v \triangleq [0,-1,0,\dots,0,-1,0]^T,
\end{align}
where $\bm v$ maps the negative sum of the odd entries onto the element $\scl[2\nx-1]$.

To minimize the required time of sensing (detailed further in Section~\ref{sec:codebook_design_num_measurements}), it is fundamental to sample the entries of $\scl$ using the minimum number of batches $M$. We propose a codebook designed to extract sequentially $\nrf\times\nrf$ contiguous, overlapping diagonal blocks of $\Sx$, resulting for $\nrf=4$ and $\nrf=2$ in the beam pattern illustrated in Fig.~\ref{fig:codebook_beampattern}. The switch configuration matrix is constructed as
\begin{align}
    \Ixm=[\bm{e}^{\nx}_{(m\Deltarf)_{\nx}},\dots,\bm{e}^{\nx}_{(\nrf-1+m\Deltarf)_{\nx}}],
\end{align}
where $\Deltarf\triangleq\nrf-1$ represents the index shift between successive batches. By employing $\Ixm$, the $m$-th batch covariance matrix $\Sm=\Ixm^T\Sx\Ixm$ extracts a sub-block of the matrix $\Sx$ such that:
\begin{align}\label{eq:Sm_uv}
    \Sm[u,v] =\Sx[(u+m\Deltarf)_{\nx},(v+m\Deltarf)_{\nx}],
\end{align}
for $0\leq u,v< \nrf$ and $0\leq m<M$. To ensure that the $2\nx$ spectral coefficients of $\scl$ are sensed, the total number of batches is chosen to satisfy\footnote{The spatial covariance matrix cannot be recovered using a single RF-chain with a fixed set of $\nx$ DFT beams, as this configuration fails to capture the inter-frequency correlation. Consequently, this restricted scenario collapses into the power-only framework analyzed in \cite{shmonin2023}.}
\begin{equation}
\label{eq:minimum_M}
M=\left\{
\begin{array}{ll}
\nx/(\nrf-1) &  1 < \nrf <\nx, \\
1 &  \nrf=\nx.
\end{array}
\right.
\end{equation}
Two key features of this codebook must be clarified:
\begin{itemize}[leftmargin=*]
\item [i)] Beam overlap. As illustrated in Fig. ~\ref{fig:codebook_beampattern} (for $\nx=4$ and $\nrf=2$), the radiation patterns associated with different beamforming matrices partially overlap, as they synthesize common DFT beams. The overlap is necessary to ensure proper sampling of the off-diagonal entries of $\Sx$. An illustrative depiction of the sampling is provided in Fig.~\ref{fig:representation_S}.
\item [ii)] Role of the $(M-1)$th beamforming matrix. Strictly speaking, the last element $\scl[2\nx-1]=\alpha_{\nx-1,0}^{-1}\cdot \Sx[\nx-1,0]$ is not required to recover $\Rx$. However, in practical scenarios where only estimates $\{\hatSm\}$ are available, it is essential to sample the entries $\Sx[0,0]$ and $\Sx[\nx-1,0]$ within the same batch. This joint measurement improves robustness in presence of errors, as evidenced by Fig.~\ref{fig:codebook_crb}, which evaluates the  Root Cramér-Rao Bound (RCRB) using each beamforming matrix $\{\Bm\}$ separately with the RCRB of the whole codebook. For this reason, the final beamforming matrix is designed to \textit{wrap around}, explicitly enforcing the simultaneous sensing of the first and last angular regions within the $(M-1)$th batch.
\end{itemize}

\subsection{ON THE NUMBER OF MEASUREMENTS}\label{sec:codebook_design_num_measurements}
The required time to estimate the DoAs, measured as the number of time snapshots  $K=M\cdot\Km$, should be minimized to enable fast target localization. The number of required snapshots is directly influenced by the nature of the transmitted signal vector $\bm s(t)$. In the ideal case where  $\bm s(t)$ is constant, the output of the equivalent FD array can be reconstructed using as few as $K=\ceil{\nx/\nrf}$ measurements \cite{berberidis2020}. However, when the transmitted signal is random, multiple observations using the same beamforming matrix are needed to build a set of estimators $\{\hat{\bm S}_m\}$, and the tight bound $K=\ceil{\nx/\nrf}$ can no longer be guaranteed. In addition, by reducing the codebook size $M$, performance can be improved by allocating more snapshots per batch.

In \cite{shmonin2023}, the authors showed that $M=2\nx$ output power measurements from a single RF chain are enough  for full covariance recovery. The algorithm can be straightforwardly extended to multi RF-chain architectures, resulting in $M=\ceil{2\nx/\nrf}$. However, methods based solely on received power discard the crucial phase relationship between the output of different RF-chains. Our proposal, in contrast, exploits the correlation between outputs and achieves accurate covariance estimation with only $M=\ceil{\nx/(\nrf-1)}$, see \eqref{eq:minimum_M}, potentially halving the codebook size. It should be noted that, by ensuring the invertibility of the estimated covariance matrices, there is an additional requirement, namely, a minimum number $K_M\geq\nrf$ of snapshots per batch. 
 However, very low numbers of  samples per batch deteriorate the performance rapidly.


\section{ULA: DERIVATION OF GLS ESTIMATOR}\label{sec:gls_derivation}
As established in Section~\ref{sec:cauchylike_driven_codebook_design}, only $2\nx-1$ unique entries from the ideal projections $\{\Sm\}$ are strictly required to perfectly recover $\Rx$, rendering any additional measurements theoretically redundant. However, in practical scenarios where only estimates $\{\hatSm\}$ are available, discarding this measurements leads to suboptimal recovery. To achieve better estimation performance, it is critical to aggregate all available information and account for the statistical characterization of the estimation errors. To achieve this, we formulate in Section~\ref{sec:codebook_spectral_representation} a linear model to map all the information contained in the set $\{\Sm\}$ onto the spatial covariance sequence $\rx$. Then, by exploiting the statistical characterization of the estimates $\{\hatSm\}$, we derive the Generalized Least Squares (GLS) estimator in Section~\ref{sec:gls_derivation_direct}. Finally, to reduce the computational complexity of a direct GLS evaluation, we propose an efficient method in Section~\ref{sec:gls_fast_implementation}, supported by a complexity analysis in Section~\ref{sec:gls_derivation_complexity}.
\subsection{CODEBOOK SPECTRAL REPRESENTATION}\label{sec:codebook_spectral_representation}
With the aim of reconstructing the SCM $\bm R_x$, now we seek  a linear model that explicitly connects the covariance matrices of the different batches, expressed in vectorized form as $\vec(\Sm)$, with the  spatial covariance sequence $\rx$. As noted previously, the covariance matrices $\{ \Sm = \Ixm^T\Sx\Ixm\}$ obtained in the different batches  extract $\nrf\times\nrf$ contiguous diagonal blocks of $\Sx$. Furthermore, we can recognize that each matrix $\Sm$ inherits the Cauchy-like displacement structure from $\Sx$ and it is fully encoded by $2\nrf-1$ real numbers embedded in the main and first off-diagonal entries:
\begin{align}\label{eq:Sm_uv2}
    &\Sm[u,v] =  \Sx[(u+m\Deltarf)_{\nx},(v+m\Deltarf)_{\nx}]\\
        &=\alphauv\cdot\begin{cases}
            \real{S_{(u+m\Deltarf)_{\nx}}-\frac{1}{\nx}S_{(u+m\Deltarf)_{\nx}}'}&u=v,\\
            \imag{S_{(u+m\Deltarf)_{\nx}}-S_{(v+m\Deltarf)_{\nx}}} & u\neq v,
        \end{cases}\nonumber
\end{align}
for $0\leq u,v<\nrf$. 
Consequently, all $\Sm$ can be expressed as a linear combination of $2\nrf-1$ basis matrices $\{\bm D_n\}_{n=0}^{2\nrf-2}$, that account for the displacement structure, weighted by a subset of entries of $\scl$. Thus, the vectorized matrix $\vec(\Sm)$ admits the following decomposition:
\begin{align}\label{eq:vec_Sm}
    \vec(\Sm) &= \sum\nolimits_{u=0}^{2\nrf-2}\scl[(2m\Deltarf+u)_{\nx}]\vec(\bm D_n)\nonumber\\
    &=\Sigmacl\Pfm\bm \Phi\bm \Omega\rx,
\end{align}
with  the involved matrices $\bm{\Omega}$ and $\bm{\Phi}$ defined in \eqref{eq:Omega} and \eqref{eq:Phi}, respectively, and 
\begin{align}
    \Sigmacl&\triangleq[\vec(\bm D_0),\dots,\vec(\bm D_{2\nrf-2})],\label{eq:Sigmacl}\\
    \Pfm&\triangleq[\bm{e}^{2\nx}_{(2m\Deltarf)_{\nx}},\dots,\bm{e}^{2\nx}_{(2(\nrf-1)+2m\Deltarf)_{\nx}}]^T.\nonumber
\end{align}
The explicit definition of the basis matrices $\{\bm D_n\}$ is provided in Appendix~\ref{appendix:A1}. Equation~\eqref{eq:vec_Sm} highlights that, in batch $m$, the spatial covariance sequence $\rx$ undergoes a cascade of linear transformations: it is first projected onto a windowed $2\nx$-DFT basis via $\bm \Phi\bm \Omega$,  then a subset of $2\nrf-1$ spectral coefficients is selected using the frequency-selection matrix $\Pfm$, and finally these coefficients are mapped to the Cauchy-like displacement structure through $\Sigmacl$.  

If we aggregate all the batch covariance matrices in a vector to make explicit the dependence with respect to the covariance sequence, such a vector $\bm p(\rx)\in\mathbb{C}^{M\nrf^2\times 1}$ admits the following compact expression:
\begin{align}\label{eq:p_rx}
    \bm p(\rx) &\triangleq  [
        \vec(\bm S_0)^T,\dots,\vec(\bm S_{M-1})^T]^T=\bm \Psi\bm \Omega\rx,
\end{align}
with 
\begin{equation}
    \bm \Psi \triangleq (\bm I_{M}\otimes \Sigmacl)\Pf\bm \Phi \;  
    \in\mathbb{C}^{M\nrf^2\times(2\nx-1)},
\end{equation}
where  
\begin{equation}\label{eq:Pf}
    \Pf\triangleq[\bm P_{f,0}^T ,\dots,\bm P_{f,M-1}^T]^T
\end{equation}
 represents the global frequency-selection matrix, formed by stacking the individual selection matrices across all batches. As illustrated in Fig.~\ref{fig:Pf}, the structure of $\Pf$ (shown for $\nx=4$ and $\nrf=2$) acts as an explicit frequency map that identifies which spectral components are sensed in each instance of the codebook. The matrix is organized into two primary blocks. The upper block captures the spectral components necessary to span the full $2\nx$-point DFT grid; its characteristic staircase structure highlights the overlap required for reconstruction, where specific DFT bins are repeatedly sensed in successive batches (see Figs. \ref{fig:codebook_beampattern} and \ref{fig:representation_S}). Conversely, the lower block accounts for the wrap-around frequencies. The specific structure of $\Pf$ is instrumental for computational efficiency,  as detailed in Section~\ref{sec:gls_derivation}.

\subsection{GLS DERIVATION}\label{sec:gls_derivation_direct}
In practical applications, the true batch covariance matrices $\{\Sm\}$ are not available, and must be estimated from a finite number of snapshots, typically via the sample average estimator:
\begin{equation}\label{eq:hat_Sm}
    \hatSm \triangleq \frac{1}{\Km}\sum\nolimits_{t=0}^{\Km-1}\bm y_{m}(t)\bm y_{m}^H(t) \;\; 0\leq m<M.
\end{equation}
Let $\hatp\in\mathbb{C}^{M\nrf^2\times 1}$ be the aggregate observation vector formed by stacking the vectorized sample average estimates. The vector $\hatp$ can be expressed as the sum of its expectation $\mathbb{E}[\hatp]=\bm p(\rx)=\bm{\Psi}\bm{\Omega}\rx$, derived in Section~\ref{sec:codebook_spectral_representation},  and a random perturbation vector $\bm{\epsilon}\in\mathbb{C}^{M\nrf^2\times 1}$ that accounts for the finite-batch effects:
\begin{align}
    \hatp &\triangleq[\vec(\hat{\bm S}_0)^T,\dots,\vec(\hat{\bm S}_{M-1})^T]^T=\bm{\Psi}\bm{\Omega}\rx+\bm{\epsilon}.
\end{align}
Using the results of \cite{ottersten1998,li1999,matiwax1999}, the vector $\bm{\epsilon}$ is asymptotically zero-mean complex Gaussian with covariance matrix equal to  
\begin{align}\label{eq:Reps}
    \Reps \triangleq 
    \frac{1}{\Km}\blkdiag(\bm S_{0}^T\otimes\bm S_{0},\dots,\bm S_{M-1}^T\otimes\bm S_{M-1}).
\end{align}
Since the error covariance matrix $\Reps$ is not a scaled identity matrix, standard Least Squares (LS) techniques \cite{li2020} yield statistically inefficient estimators. To achieve asymptotic efficiency, we leverage the GLS framework by incorporating the error statistics into the following weighted minimization problem:
\begin{equation}\label{eq:opt_fgls}
    \hatrx = \arg \min_{\rx}\|\hatReps^{-1/2}(\bm p(\rx)-\hatp)\|_2^2.
\end{equation}
This formulation is technically referred to as a \textit{Feasible Generalized Least Squares} estimator because the true covariance $\Reps$ is replaced by the consistent estimate $\hatReps$, constructed by using the sample averages estimates $\hatSm$ \eqref{eq:hat_Sm} in the covariance matrix \eqref{eq:Reps} 
instead of the true covariance matrices $\Sm$. 
As $\bm \Omega$ in \eqref{eq:Omega} is invertible (see Appendix \ref{appendix:A1}), the optimization problem \eqref{eq:opt_fgls} admits a closed-form solution:
\begin{align}\label{eq:hatrx}
    \hatrx &= \bm \Omega^{-1}(\bm \Psi^H\hatReps^{-1}\bm \Psi)^{-1}\bm \Psi^H\hatReps^{-1}\hatp.
\end{align}
The high complexity of this computation can be significantly reduced, as we see next. 


The existence and uniqueness
of the GLS estimator $\hatrx$ hinges on the non-singularity of all $\hatReps$, $\bm \Omega$, and  $(\bm \Psi^H\hatReps^{-1}\bm\Psi)$. If the number of samples per batch satisfies $\Km\geq \nrf$, the sample covariance estimates $\{\hatSm\}$ in \eqref{eq:hat_Sm} are positive definite \cite{trees2002}, ensuring $\hatReps$ is invertible. Moreover, the invertibility of the square matrix $\bm \Omega$ is demonstrated in Appendix~\ref{appendix:A2}. Consequently, the existence of $\hatrx$ reduces to requiring that $\bm \Psi$ be full column rank, i.e., $\rank(\bm \Psi)=2\nx-1$. This condition is established in the following lemma.
\begin{lemma}\label{lemma:ula} If the codebook size satisfies $M\geq \ceil{\nx/(\nrf-1)}$, then the matrix $\bm \Psi$ has full column rank, namely $\rank(\bm \Psi)=2\nx-1.$ \end{lemma}
\begin{proof} We rely on a property of matrix ranks: for two matrices $\bm A$ and $\bm B$, if $\bm A$ has full column rank, the rank of their product satisfies $\rank(\bm A\bm B)=\rank(\bm B)$ \cite{bernstein2009}. Using this argument, we can evaluate the rank of the linear operator $\bm \Psi=(\bm I_M\otimes\Sigmacl)\Pf\bm \Phi$ by sequentially checking that its constituent matrices possess full column rank from left to right: \textit{i)} The rank of the Kronecker product is the product of the rank of its operands: $\rank(\bm I_M\otimes\Sigmacl)=\rank(\bm I_M)\cdot\rank(\Sigmacl)$. Since $\Sigmacl$ is shown to have column rank in Appendix~\ref{appendix:A1}, the matrix $(\bm I_M\otimes\Sigmacl)$ has full column rank. \textit{ii)} By choosing a codebook size  with $M\geq\ceil{\nx/(\nrf-1)}$, we guarantee that $\Pf$ defined in \eqref{eq:Pf} contains the identity matrix $\bm I_{2\nx-1}$ as a sub-matrix, which directly implies it has full column rank. \textit{iii)} Analogously to the previous step, the operator $\bm \Phi$ defined in \eqref{eq:Phi} contains the identity matrix $\bm I_{2\nx-1}$ as a sub-matrix.
Following this sequential process, we conclude that $\rank(\bm \Psi) = 2\nx-1$. \end{proof}

\begin{algorithm}[H]
\caption{Fast GLS Estimator}
\label{alg:pseudocode_fast_solver} 
\begin{algorithmic}[1]
    \Require $M, \{\hat{\bm{S}}_m\}_{m=0}^{M-1}, \bm{P}_f, \bm{\Phi}, \bm \Omega, \Km$
    
    \Statex \rule{\linewidth}{0.4pt}
    \Statex \centerline{\textbf{Stage I: Batch Cauchy-like processing}} \vspace{-5pt}
    
    \For{$m = 0$ \textbf{to} $M-1$}
        \State Compute $\Sigmacl^H\vec(\hatSm^{-1})$ using \eqref{eq:trace0}
        \State Compute $\Sigmacl^H\hatRepsm^{-1}\Sigmacl$ using \eqref{eq:trace1}
    \EndFor
    \State  Compute $(\bm I_M\otimes\Sigmacl)^H\hatReps^{-1}\hatp$ via \eqref{eq:block_mat_vec}
    \State  Compute $(\bm I_M\otimes\Sigmacl)^H\hatReps^{-1}(\bm I_M\otimes\Sigmacl)$ via \eqref{eq:block_mat_mat}

    \Statex \rule{\linewidth}{0.4pt}
    \Statex \centerline{\textbf{Stage II: Banded and Low-rank Decomposition}} \vspace{-5pt}
    
    \State Form block matrices $\bm M_{00}, \bm M_{01}, \bm M_{10}, \bm M_{11}$ via \eqref{eq:M_ij}
    \State Construct $\bm Q, \bm U,$ and $\bm C$ via \eqref{eq:Q}, \eqref{eq:U}, \eqref{eq:C}

    \Statex \rule{\linewidth}{0.4pt}
    \Statex \vspace{0.2cm} \centerline{\parbox{8cm}{\centering\textbf{Stage III: \\Estimation of $\hatscl$ via Woodbury Inversion Lemma}}}
    
        \State  $\bm V \gets \bm C^{-1}+\bm U^H\bm Q^{-1}\bm U$ (banded solver)
        \State  $\bm z_0 \gets \bm \Phi^T\Pf^T[(\bm I_M\otimes\Sigmacl^H)\hatReps^{-1}\hatp]$
        \State  $\bm z_1 \gets \bm V^{-1}\bm U^H\bm z_0$
        \State  $\hatscl\gets \bm z_0-\bm Q^{-1}\bm U\bm z_1$ (banded solver)

    \Statex \rule{\linewidth}{0.4pt}
    \Statex \centerline{\textbf{Stage IV: Inverse Fast Fourier Transform}} \vspace{-5pt}
    
    \State $\hat{\bm X} \gets \text{IFFT}[\hatscl]$
    \State $\hatrx \gets \hat{\bm X}$ by solving $\nx$ $2\times 2$ linear systems \eqref{eq:decoupled_systems}
\end{algorithmic}
\end{algorithm}

\subsection{GLS FAST IMPLEMENTATION}\label{sec:gls_fast_implementation}
Direct evaluation of the GLS estimator \eqref{eq:hatrx} entails significant computational overhead, stemming primarily from the high-dimensional matrix product $\bm \Psi^H\hatReps^{-1}\bm \Psi$ and the subsequent inversion. Specifically, given a codebook size of $M=\ceil{\nx/(\nrf-1)}$, the complexity of the product scales as $\mathcal{O}(\nx(M\nrf)^2)=\mathcal{O}(\nrf^2\nx^3)$, while the inversion requires $\mathcal{O}(\nx^3)$. However, the block-diagonal nature of $\hatReps^{-1}$ and the algebraic structure of $\bm \Psi=(\bm I_M\otimes\Sigmacl)\Pf\bm\Phi$ permits a more efficient implementation where the direct evaluation of $\bm \Psi^H\hatReps^{-1}\bm\Psi$ can be bypassed by leveraging the  structure of the Cauchy-like operator $\Sigmacl$. 
Furthermore, by recognizing that $\bm \Psi^H\hatReps^{-1}\bm \Psi$ can be expressed as a low-rank update of a banded matrix\footnote{An arbitray matrix $\bm A$ is said to be banded with bandwidth $b$ if $\bm A[u,v]=0$ for $|u-v|>b$.}, the direct inversion is avoided entirely, reducing the overall complexity to $\mathcal{O}(\nrf^2\nx)$. Once these operations are simplified, the remaining bottleneck is the multiplication by $\bm \Omega^{-1}$; however, the DFT-like structure of $\bm \Omega$ in \eqref{eq:Omega} enables the use of the Inverse Fast Fourier Transform (IFFT) to circumvent this multiplication.

We partition the fast implementation into four stages. For readability purposes, we summarize the main steps of each stage below, with comprehensive mathematical derivations provided in Appendix~\ref{appendix:B}. Furthermore, Algorithm~\ref{alg:pseudocode_fast_solver} presents the corresponding pseudocode.
\begin{itemize}[leftmargin=*]
    \item \textit{Stage I: Batch Cauchy-like processing (\ref{appendix:B1})}. This stage involves the inversion of  $M$ matrices $\{\hatSm^{-1}\}$, and the evaluation of the corresponding matrix-vector and matrix-matrix products, $\{\Sigmacl^H(\hatSm^{-T}\otimes\hatSm^{-1})\vec(\hatSm)\}$ and $\{\Sigmacl^H(\hatSm^{-T}\otimes\hatSm^{-1})\Sigmacl\}$, respectively. While the inversion of dense unstructured matrices $\{\hatSm\}$ offers no savings, the associated products are streamlined by exploiting the Cauchy-like displacement structure encoded in $\Sigmacl$. Upon completion, the terms $(\bm I_M\otimes\Sigmacl^H)\hatReps^{-1}(\bm I_M\otimes\Sigmacl)$ and $(\bm I_M\otimes\Sigmacl^H)\hatReps^{-1}\hatp$ are available for use in the subsequent phases.
    \item \textit{Stage II: Banded and Low-rank Decomposition (\ref{appendix:B2})}. We exploit the decomposition  $\bm \Psi^H\hatReps^{-1}\bm \Psi=\bm Q+\bm U\bm C\bm U^H$. Here, $\bm Q$ is a banded matrix with bandwidth $2\nrf-1$, a structure arising from the staircase-like pattern of $\Pf$ (see Fig.~\ref{fig:Pf}). The low-rank component $\bm U\bm C\bm U^H$ accounts for the effect of $\bm \Phi$ and the wrap-around of $\Pf$.
    \item \textit{Stage III: Estimation of $\hatscl$ via Woodbury Inversion Lemma (\ref{appendix:B3})}. This stage derives the estimate of the spectral vector $\hatscl$. Based on the linear correspondence $\scl=\bm \Phi\bm \Omega\rx$ established in \eqref{eq:scl_mat}, the GLS estimate of $\scl$ is defined as:
\begin{equation}\label{eq:hatscl}
    \hatscl\triangleq \bm \Phi \bm \Omega \hatrx=\bm \Phi (\bm \Psi^H\hatReps^{-1}\bm \Psi)^{-1}\bm \Psi^H\hatReps^{-1}\hatp.
\end{equation}
Leveraging the sparse structure of $\bm \Phi$ in \eqref{eq:Phi}, the evaluation $\bm \Psi^H\hatReps^{-1}\hatp=\bm \Phi^T\Pf^T(\bm I_M\otimes\Sigmacl^H)\hatReps^{-1}\hatp$ is computationally negligible using  Stage I results. The main computational burden lies in evaluating the matrix inverse. However, explicit inversion can be avoided by applying the Woodbury Inversion Lemma together with banded linear solvers, enabling the efficient computation of $\hatscl$ without directly forming the inverse matrix.
\item \textit{Stage IV: Inverse Fast Fourier Transform (\ref{appendix:B4})}. The final stage, transform the spectral vector $\hatscl$ into the spatial covariance sequence $\hatrx$. While, $\hatrx$ could be recovered from \eqref{eq:hatscl} via a pseudo-inverse, the structure of the involved matrices makes it possible to obtain $\hatrx$ efficiently by means of an  IFFT.
\end{itemize}

\begin{table}[ht]\label{table:table_complexity}
\caption{Theoretical complexity.}
\centering
\renewcommand{\arraystretch}{1} 
\begin{tabular}{c c c} 
\hline
\textbf{Operation} & \textbf{Times} & \textbf{Complexity per operation}\\ \hline
$\hatSm^{-1}$ & $M$ & $\mathcal{O}(\nrf^3)$ \\ \hline
$\Sigmacl^H\vec(\hatSm^{-1})$ & $M$ & $\mathcal{O}(\nrf^2)$ \\ \hline
$\Sigmacl^H(\hatSm^{-T}\otimes\hatSm^{-1})\Sigmacl$ & $M$ & $\mathcal{O}(\nrf^3)$ \\ \hline
\text{Eval. }$\hatscl$ & 1 & $\mathcal{O}(\nrf^2 \nx)$ \\ \hline
\text{Eval. }$\hatrx=\bm \Omega^{-1}\hatscl$ & 1 & $\mathcal{O}(\nx\log \nx)$ \\ \hline
\end{tabular}
\label{table:theoretical_time_complexity}
\end{table}
\subsection{COMPUTATIONAL COMPLEXITY ANALYSIS}\label{sec:gls_derivation_complexity}
The computational requirements for the various stages introduced in Section~\ref{sec:gls_fast_implementation} are summarized in Table~\ref{table:theoretical_time_complexity}. Thus, for a codebook size selected according to $M=\ceil{\nx/(\nrf-1)}$, as introduced in \eqref{eq:minimum_M}, the computational complexity of $M$ inverses $\hatSm$ and the products $\Sigmacl^H(\hatSm^{-T}\otimes\hatSm^{-1})\Sigmacl$ yields
\begin{equation}
    \mathcal{O}(2M\nrf^3)=\mathcal{O}\Big(\ceil{\tfrac{\nx}{\nrf-1}}\nrf^3\Big)=\mathcal{O}(\nrf^2\nx).
\end{equation}
Thus, the complexity is primarily driven by the terms $\mathcal{O}(\nrf^2\nx)$ and $\mathcal{O}(\nx\log\nx)$ with total order of $\mathcal{O}(\max\{\nrf^2\nx,\nx\log\nx\})$. However, for moderate values of $\nrf$, the term $\mathcal{O}(\nrf^2\nx)$ dominates the computational budget. While the $\mathcal{O}(\nx\log\nx)$ associated with the IFFT theoretically dominates as the number of antennas $\nx$ grows towards infinity (for a fixed $\nrf$), its logarithmic growth rate is sufficiently slow so that $\mathcal{O}(\nrf^2\nx)$ remains the effective complexity for most practical scenarios. It is important to mention that the exploitation of the displacement structures in Stage I is what enables this efficiency. Without leveraging the structural properties, the complexity of these operations would escalate to $\mathcal{O}(\nrf^4\nx)$, effectively becoming the main computational bottleneck.

\begin{figure}[t!]
    \centering
    \includegraphics[width=1\linewidth]{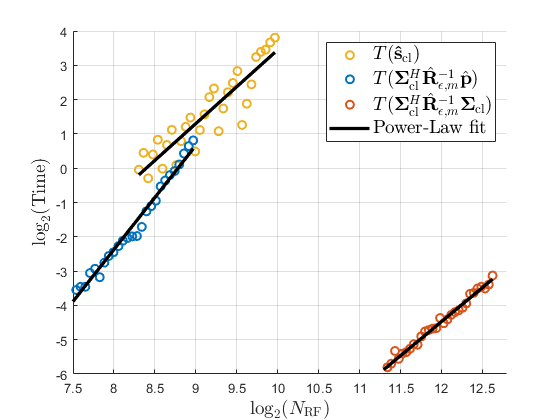}
    \caption{Evaluation of empirical complexity by means of execution time, $\nx=2000$. The  entries of $\hatRepsm$ are drawn from a complex Gaussian distribution.}
    \label{fig:empirical_time_complexity}
\end{figure}

To provide empirical validation of our theoretical analysis, we adopt the power-law model for execution time \cite{goldsmith2007measuring}. Let $T(f(x))$ denote the empirically measured execution time of an operation $f(x)$. Assuming an asymptotic complexity of $\mathcal{O}(a\cdot\nrf^b)$, 
the relationship can be linearized as:
\begin{equation}
    \log_2(T(f(x)) = \log_2(a) + b\cdot\log_2(\nrf),
\end{equation}
where $a$ is hardware dependent constant specific to the simulation environment, and $b$ is the scaling exponent to be estimated. We characterized $b$ empirically by measuring the execution time using the MATLAB\textsuperscript{\textregistered} \textsc{Timeit} function. For the evaluation of the empirical complexity the entries of the Hermitian matrices $\{\hatRepsm\}$ were generated
randomly from a complex Gaussian distribution. The  results, summarized in Table~\ref{table:empirical_time_complexity} and represented in Fig.~\ref{fig:empirical_time_complexity}, demonstrate a close agreement between the theoretical exponent and our empirical measurements. 
Notably, Fig.~\ref{fig:empirical_time_complexity} reveals a higher dispersion for $T(\hatscl)$ compared to the other operations. This variance stems from the fact that the number of operations required for computing $\hatscl$ depends significantly on the dimensions of the wrap-around block, which fluctuates according to the $\nx/\nrf$ ratio.

\begin{table}[t!]
\caption{Empirical time complexity.}
\centering
\renewcommand{\arraystretch}{1} 
\begin{tabular}{c c c} 
\hline
\textbf{Operation} & \textbf{Theoretical b} & \textbf{Estimated b} \\ \hline
$\Sigmacl^H\vec{\hatSm^{-1}}$  & $2$ & $1.9849$  \\ \hline
$\Sigmacl^H(\Sm^{-T}\otimes\hatSm^{-1})\Sigmacl$  & $3$ & $3.0323$ \\ \hline
$\hatscl$  & $2$ & $2.1491$ \\ \hline
\end{tabular}
\label{table:empirical_time_complexity}
\end{table}
\section{URA: SIGNAL MODEL}
\begin{figure}[t!]
    \centering
\includegraphics[width=1\linewidth]{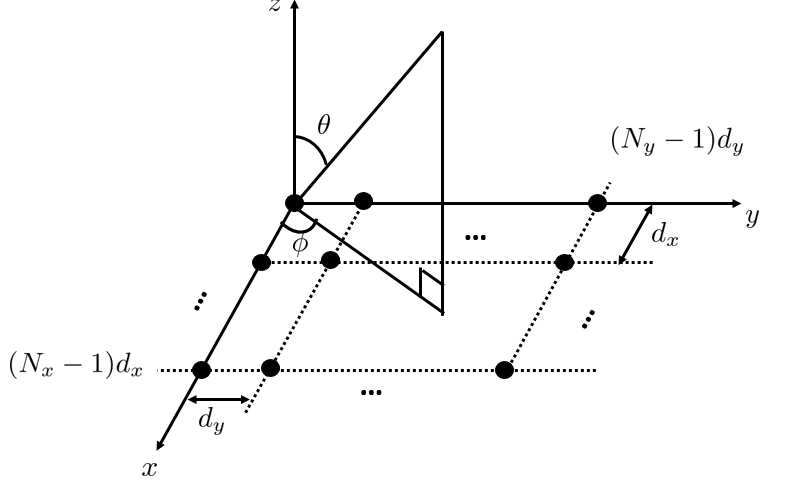}
    \caption{Geometry of Uniform Rectangular Array.}
    \label{fig:ura_array}
\end{figure}

Building upon the framework established for ULAs, this section generalizes the proposed method to two-dimensional uniform rectangular arrays (URAs). The geometry of the array is shown in Fig.~\ref{fig:ura_array}, where $N=\nx\ny$ antennas are uniformly arranged over a rectangular grid at positions $\{(u\cdot d_x,v\cdot  d_x)\}$ with $0\leq u<\nx$ and $0\leq v<\ny$. We assume half-wavelength spacing again  $d_x=d_y=\lambda/2$.

In this architecture, the antennas are also connected to a passive processing structure, such as a two-dimensional Butler Matrix  \cite{ding2018,cavalcante2022efficient}, which performs analog beamforming via the 2D-DFT matrix 
\begin{equation}
    \bm{F}\triangleq\bm{F}_x\otimes\bm{F}_y,
\end{equation}
where $\bm{F}_x\in\mathbb{C}^{\nx\times\nx}$ and $\bm{F}_y\in\mathbb{C}^{\ny\times\ny}$ are the DFT matrices defined in \eqref{eq:DFT}. Subsequently, a switching network selects $\nrf$ outputs for down-conversion and sampling. 

As in the ULA case, the received signal is collected into 
$M$ batches, with a fixed switch configuration applied throughout each batch over $\Km$  time instants. 
To characterize the network, the number of RF-chains is factored as $\nrf=\nrfx\cdot\nrfy$ (with $\nrfx,\nrfy>1$), and the $m$th switch configuration is defined as 
\begin{equation}
    \Im\triangleq\Ixm\otimes\Iym.
\end{equation}
 Here, $\Ixm\in\{0,1\}^{\nx\times\nrfx}$ and $\Iym\in\{0,1\}^{\ny\times\nrfy}$ control the selection of vertical and horizontal Butler Matrix outputs, respectively. After down-conversion and sampling, the received signal yields
\begin{equation}
    \bm y_m(t) = \bm B_m^H(\bm A(\bm \psi)\bm s(t+mK_M) + \bm n(t+mK_M)),
\end{equation}
with $0\leq m < M$ and  $0\leq t < \Km-1$. The random vector $\bm s(t)=[\bm s_0(t),\dots,s_{L-1}(t)]^T\in\mathbb{C}^{L\times 1}$ contains the source signals and  $\bm n(t)\in\mathbb{C}^{N\times 1}$ represents the additive noise. Both vectors follow the same statistical characterization as in the ULA scenario discussed earlier. 
The beamforming matrix $\Bm\in\mathbb{C}^{N\times\nrf}$ is composed of columns from the 2D-DFT matrix $\bm F$ as
\begin{equation}
\bm B_m = \bm F\Im = (\Fx\otimes\Ixm)(\Fy\otimes\Iym),    
\end{equation} 
where we are using the Kronecker product property
$(\Fx\otimes\Fy)\cdot(\Ixm\otimes\Iym)=(\Fx\otimes\Ixm)(\Fy\otimes\Iym)$. The two-dimensional array manifold is defined as $\bm A(\bm \psi)=[\bm a(\bm \psi_{0}),\dots,\bm a(\bm \psi_{L-1})]$, where the vector $\bm{\psi}_{\ell}=[\psi_{x,\ell},\psi_{y,\ell}]^T$ encodes the angular information of the $\ell$th source as $\bm \psi_{\ell}\triangleq [\pi\sin\theta_{\ell}\cos\phi_{\ell}, \;\pi\sin\theta_{\ell}\sin\phi_{\ell}]^T$
 with $\theta_\ell$ and $\phi_\ell$ denoting the elevation and azimuth angles, respectively.
 Assuming perfect calibration, the steering vector of the $\ell$-th source $\bm a(\bm{\psi}_\ell)\in\mathbb{C}^{N\times 1}$ is expressed as the Kronecker product of two ULA steering vectors with $\nx$ and $\ny$ antennas, respectively:
\begin{equation}
    \bm a(\bm \psi_\ell) \triangleq \bm a_x(\psi_{x,\ell})\otimes\bm a_y(\psi_{y,\ell}).
\end{equation}
Under these assumptions, the SCM $\bm R\in\mathbb{C}^{N\times N}$ of the FD array is given by 
\begin{align}
    \bm R =\sum\nolimits_{\ell=0}^{L}(\Rxl\otimes\Ryl),
\end{align}
where we incorporate the powers into the matrices as $\Rxl=\sigma^2_{s,\ell}\bm  a(\psixl)\bm a^H(\psixl)$ and $\Ryl=\bm  a_y(\psiyl)\bm a_y^H(\psiyl)$ for $0\leq \ell <L$, and $\bm R_{x,L}=\sigma_n^2\bm I_{\nx}$ and $\bm R_{y,L}=\bm I_{\ny}$. The matrices of the sets $\{\Rxl\}$ and $\{\Ryl\}$ are Hermitian Toeplitz, so they are defined by the spatial sequences $\rxl\in\mathbb{C}^{(2\nx-1)\times 1}$ and $\ryl\in\mathbb{C}^{(2\ny-1)\times 1}$, respectively. Consequently, $\bm R$ follows a Block Toeplitz with Toeplitz Blocks (BTTB) structure \cite{heidenreich2011} and can be characterized by only $(2\nx-1)\cdot(2\ny-1)$ unique entries.

Analogous to the ULA case, the batch covariance matrices are samples of the transformed covariance matrix $\bm S\triangleq\bm F^H\bm R\bm F$. Specifically, the $m$th batch covariance matrix satisfies
\begin{equation}\label{eq:ura_Sm}
    \Sm =  \sum\nolimits_{\ell=0}^{L}\Sxlm\otimes \Sylm.
\end{equation}
Unlike the ULA scenario, $\bm S$ does not possess a Cauchy-like displacement structure. However, the constituent matrices $\Sxlm=\Ixm^T\Fx^H\Rxl\Fx\Ixm$ and $\Sylm=\Iym^T\Fy^H\Ryl\Fy\Iym$ retain the same structure identified in the ULA. We exploit this property in the following section to formulate the linear measurement model for the GLS estimator.

\section{URA: DERIVATION OF GLS ESTIMATOR}
This section introduces the GLS estimator for URA arrays and proves its mathematical validity under the proposed codebook design. The remainder of this section is organized as follows. First, Section~\ref{sec:ura_codebook_design}, introduces a codebook architecture for URAs as a natural extension of the 1D ULA framework. Section~\ref{sec:ura_codebook_spectral_representation} establishes a linear measurement model by leveraging the Cauchy-like displacement structure of the covariance matrix batch components $\{\Sxlm\}$ and $\{\Sylm\}$ outlined in  \eqref{eq:ura_Sm}. Finally, Section~\ref{ura:gls_derivation} details the formulation of the GLS estimator and, similarly to the  ULA derivation, demonstrates the existence and uniqueness of the estimator.
\subsection{URA: CODEBOOK DESIGN}\label{sec:ura_codebook_design}
The proposed codebook is a natural extension of the ULA design to the URA scenario. Specifically, the total number of beamforming matrices is given by:
\begin{align}
    M = \mx\cdot \my
\end{align}
where $M_{x/y}$ are defined as the codebook size in ULAs \eqref{eq:minimum_M} using $N_{x/y}$ antennas and $N_{\text{RF},x/y}$ RF-chains. Using these parameters, the $m$th selection matrix $\Im$ is constructed via
\begin{equation}\label{eq:ura_Im}
    \Im = \bm{\mathcal{I}}_{x,\floor{m/\my}} \otimes \bm{\mathcal{I}}_{y,(m)_{\my}} \quad 0\leq m<M.
\end{equation}
Intuitively, while the ULA codebook was designed to sample the windowed spectral density accross the $\nx$-DFT beams and the inter-frequency correlation between adjacent beams, the URA codebook extends this paradigm to account for both spatial dimensions. 


Analogously to the ULA case, our design assumes $\nrfx > 1$ and $\nrfy > 1$ to guarantee access to the inter-frequency correlations between adjacent beams. Scenarios where the number of RF chains along one or both dimensions equals one require a hybrid approach. Specifically, one would need to integrate the power-based method proposed in \cite{shmonin2023} for the single-RF case with the framework derived in the following section.

\subsection{URA: CODEBOOK SPECTRAL REPRESENTATION}\label{sec:ura_codebook_spectral_representation}
The $m$th batch covariance matrix $\Sm$ in \eqref{eq:ura_Sm} comprises a sum of $L$ Kronecker products of Cauchy-like matrices. This structural property enables the generalization of the linear measurement model of ULAs, derived in Section~\ref{sec:codebook_spectral_representation}, to the URA scenario. In order to formalize this, we invoke the relationship between the vectorization operator and the Kronecker product. For any two matrices $\bm A\in\mathbb{C}^{r\times s}$ and $\bm B\in\mathbb{C}^{p\times q}$, the vectorization of their Kronecker product is given by
$\vec(\bm A \otimes \bm B) = ( \bm I_s \otimes \bm{K}_{q,r} \otimes \bm I_p)(\vec(\bm A) \otimes \vec(\bm B))$, where $\bm K_{q,r} \in \{0,1\}^{qr \times qr}$ is the \textit{commutation matrix} \cite{bernstein2009} , which serves as the unique permutation matrix that reorders the elements of a vectorized matrix to match its transpose, i.e., $\bm K_{q,r}\vec(\bm C) = \vec(\bm C^T)$. 

Thus, by defining the operator 
\begin{equation}
    \bm J_{xy} \triangleq ( \bm I_{\nrfx} \otimes \bm{K}_{\nrfy,\nrfx}\otimes \bm I_{\nrfy}),
\end{equation}
we can express the vectorization of the $m$th batch covariance matrix, $\vec(\Sm)$, by leveraging the 1D spectral models derived for the horizontal and vertical components as
\begin{equation}
    \vec(\Sm)=\bm{J}_{xy}\sum\nolimits_{\ell=0}^{L}(\vec(\Sxlm)\otimes\vec(\Sylm))\nonumber.
\end{equation}
For convenience, we will use  $\bm \Phi_{x/y}$, $\bm \Omega_{x/y}$, $\bm{\Sigma}_{\text{cl},{x/y}}$ and $\bm P_{f,m,x/y}$ to denote the application of the corresponding  ULA matrices to the parameters $(\nx,\nrfx)$ and $(\ny,\nrfy)$, respectively. With this, the vectorized matrix $\vec(\Sm)$ yields
\begin{align}\label{eq:ura_vec_Sm}
    &\vec(\Sm) = \bm{J}_{xy}\Sigmaclxy\Pfmxy\Phixy\Omegaxy\rxy
\end{align}
where the operators are formed as the Kronecker products of their 1D counterparts: 
$\Sigmaclxy \triangleq (\Sigmaclx\otimes\Sigmacly)$, $\Pfmxy \triangleq (\Pfmx\otimes\Pfmy)$, $\Omegaxy = (\Omegax\otimes\Omegay)$ and $\Phixy\triangleq(\bm \Phi_x\otimes\bm \Phi_y)$.
We also define the vector
\begin{equation}
    \rxy \triangleq  \sum\nolimits_{\ell=0}^{L}(\rxl\otimes\rxl)\in\mathbb{C}^{(2\nx-1)(2\ny-1)\times 1}
\end{equation}
as the spatial covariance sequence of the URA array. 

As shown in \eqref{eq:ura_vec_Sm}, the URA measurement model maintains a structural form analogous to the ULA scenario \eqref{eq:vec_Sm}. However, while the ULA utilizes a 1D-DFT to transform the spatial covariance into the frequency domain, the URA employs the composite operator $(\Phix\otimes\Phiy)(\Omegax\otimes\Omegay)$ to project $\rxy$ onto a 2D grid of $2\nx\times 2\ny$ DFT points. In this 2D spectral plane, the matrix $\Pfmxy=\Pfmx\otimes\Pfmy$ acts as a frequency selection mask, by sampling specific combinations of horizontal and vertical spatial frequencies.

Moreover, under the codebook design introduced in Section~\ref{sec:ura_codebook_design}, stacking the $M=\mx\cdot\my$ individual selection matrices as $\Pfxy=[\bm P_{f,0,xy}^T,\dots,\bm P_{f,M-1,,xy}^T]^T$, the matrix $\Pf$ satisfies
\begin{align}
        &\Pfxy = [ \bm P_{f,0,x}^T\otimes\bm P_{f,0,y}^T, \ (\bm P_{f,0,x}^T\otimes\bm P_{f,1,y}^T, \ \dots \nonumber \\
&\dots, \ \bm P_{f,\mx-1,x}^T\otimes\bm P_{f,\my-1,y}^T]^T=\bm{\mathcal{P}}(\Pfx\otimes\Pfy),\label{eq:ura_Pf}
\end{align}
where $\bm{\mathcal{P}}$ is a permutation matrix that accounts for the row reordering. The matrices $\Pfx\triangleq[\bm P_{f,0,x}^T,\dots,\bm P_{f,\mx-1,x}^T]^T$ and $\Pfy\triangleq [\bm P_{f,0,y}^T,\dots,\bm P_{f,\my-1,y}^T]^T$ are the spectral selectors derived for ULAs in \eqref{eq:Pf}, constructed for codebook sizes  $\mx$ and $\my$, respectively.

\subsection{URA: GLS DERIVATION}\label{ura:gls_derivation}
Once more, practical access to the difference batch covariance matrices $\{\hatSm\}$ is obtained via sample average over $\Km$ snapshots. To estimate the 2D spatial covariance sequence $\hatrxy$, we formulate a Generalized Least Squares (GLS) optimization problem. By aggregating all $M$ batch estimates into the vector $\hatp=[\vec(\hat{\bm S}_0)^T,\dots,\vec(\hat{\bm S}_{M-1})^T]^T$, the objective function is defined as:
\begin{equation}
\hatrxy = \arg \min_{\rxy}
\|\hatReps^{-1/2}(\hatp-\Psixy\Omegaxy\rxy)\|_{2}^{2},
\end{equation}
where $\hatReps$ is the estimated error covariance matrix, computed analogously to the ULA case in \eqref{eq:Reps}, and the operator $\Psixy$ is defined as $\Psixy\triangleq (\bm I_M\otimes\Sigmaclxy)\Pfxy\Phixy$. The closed-form solution for the reconstructed 2D spatial covariance sequence is given by:
\begin{equation}\label{eq:hatrxy}
    \hatrxy = \Omegaxy^{-1}(\Psixy^H\hatReps^{-1}\Psixy)\Psixy^H\hatReps^{-1}\hatp.
\end{equation}
Unlike the ULA scenario, we cannot easily leverage the specific structure of the operator $\Psixy$ to further reduce the computational overhead. While the constituent matrices $\Pfxy$ and $\Phixy$ are extremely sparse, the solution in \eqref{eq:hatrxy} requires the inversion of a matrix of size $(2\nx-1)(2\ny-1)$. For moderate array sizes, this inversion is computationally manageable; however, it becomes prohibitive as the number of antennas increases. In such high-dimensional regimes, a practical alternative involves estimating the DoAs for the horizontal and vertical dimensions separately. By applying a data association step to pair the resulting angles, one can significantly reduce the overall complexity. While this decoupled approach is sub-optimal compared to the joint GLS estimator, it offers a necessary trade-off for real-time feasibility in large-scale URAs \cite{trees2002}.

Assuming the number of snapshots per batch satisfies $\km\geq \nrf$, the matrix $\hatReps$ is guaranteed to be invertible with probability one. Furthermore,  matrix $\Omegaxy$ is invertible since the Kronecker product preserves non-singularity yielding $\Omegaxy^{-1}=\Omegax^{-1}\otimes\Omegay^{-1}$, where the constituent matrices are full rank as demonstrated in Appendix~\ref{appendix:A2}. Consequently, the uniqueness of $\hatrxy$ depends solely on the full column rank of $\Psixy$, guaranteed by the following lemma.
\begin{lemma}
    A codebook with $M\geq \ceil{\nx/(\nrfx-1)}\cdot \ceil{\ny/(\nrfy-1)}$ beamforming matrices guarantees a full column rank  $\Psixy$, i.e.,  $\rank(\Psixy)=(2\nx-1)\cdot(2\ny-1)$.
\end{lemma}
\begin{proof}Following the arguments in Lemma~\ref{lemma:ula} for the ULA, we prove that the matrix $\Psixy=\bm J_{xy}(\bm I_M\otimes\Sigmaclxy)\Pfxy\Phixy$ has full column rank. The commutation matrix $\bm J_{xy}$ is full rank by definition. Applying Kronecker rank property $\rank(\bm A\otimes\bm B) = \rank(\bm A)\cdot\rank(\bm B)$ alongside with the results of Lemma~\ref{lemma:ula}, the factors $\bm I_M\otimes\Sigmaclxy=\bm I_M\otimes\Sigmaclx\otimes\Sigmacly$ and $\Phixy=(\Phix\otimes\Phiy)$ possess full column rank. Finally, since multiplication by the permutation matrix $\bm{\mathcal{P}}$ preserves rank, the component $\Pfxy=\bm{\mathcal{P}}(\Pfx\otimes\Pfy)$ also has full column rank. Consequently, $\Psixy$ satisfies $\rank(\Psixy)=(2\nx-1)(2\ny-1)$.\end{proof}


\section{SIMULATIONS}


We simulate an array with inter-element spacing $d=\tfrac{\lambda}{2}$ under narrowband signal conditions. To simplify the analysis, all source signals are assumed to have equal power normalized to one. Therefore, the signal-to-noise ratio (SNR) is computed as $\text{SNR} \triangleq -10\log_{10}\sigma_n^2$. Once the covariance matrix is estimated, the Direction-of-Arrivals (DoAs) are obtained using Root-MUSIC \cite{trees2002}. The main performance metric of the bench marked methods is the Root Mean Squared Error (RMSE), defined as
\begin{equation}\label{eq:RMSE}
    \text{RMSE} \triangleq \sqrt{\frac{1}{L\cdot \text{MC}}\sum_{i=0}^{\text{MC}-1}\sum_{l=0}^{L-1}(\theta_{l}-\hat{\theta}_{i,l})^2}
\end{equation}
with $L$  the number of signals, $\text{MC}$ the number of Monte Carlo realizations, and $\{\theta_{l},\hat{\theta}_{i,l}\}$  the true and estimated angles of arrival, respectively. The number of Monte Carlo realizations is fixed for all experiments to $\text{MC}=10^4$. We evaluate performance using the Root Cramér-Rao Lower Bound (RCRB), defined as $\text{RCRB}\triangleq \sqrt{\tr{\text{CRB}(\bm \theta)}/L}$, where the expression for $\text{CRB}(\bm \theta)$ can be found in \cite{matiwax1999}.

Our proposal, denoted Cauchy-like GLS (CL-GLS) is evaluated against three existing covariance estimation approaches. Our earlier 2D-DFT Least Squares (2D-DFT LS) \cite{rivas2025}, which employs the same codebook but neglects error statistics. The Beam Sweeping Algorithm (BSA) \cite{li2020}, and the Phase-Independent (PI) recovery \cite{shmonin2023} are power-based method. Both use $2\nx$ DFT beams distributed across a total of $M=\ceil{2\nx/\nrf}$ beamforming matrices, with the design of the DFT beams per batch following the strategy proposed in \cite{li2020}. We also  included Dynamical Covariance Orthogonal Matching Pursuit (DCOMP) \cite{Park2018} as a reference of Compressed Sensing methods, where the codebook consists of $K$ different beamforming matrices with pseudo-random beamforming\footnote{In pseudo-random beamforming the phase of each beamforming matrix is drawn from a uniform distribution i.e. $\angle(\bm B_m[u,v])\sim\mathcal{U}(0,2\pi)$, $\forall u,v,m$.}. Finally, we benchmark our proposal against the method in \cite{matiwax1999}, hereafter denoted as Exhaustive Asymptotic Maximum-Likelihood (E-AML).  Although E-AML also employs a GLS objective to approximate the ML estimator, it targets the direct estimation of DoA parameters rather than the SCM itself. This formulation introduces several limitations: it requires a search over an $L$-dimensional cost function increasing the complexity as the number of sources grows. The authors recommend the use of Alternate Projections to reduce the overall complexity but, even in that scenario, the required complexity is very high. Additionally, although targeting directly the DoAs may provide better accuracy, E-AML discards potential useful information encoded in the SCM that can be used, for example to estimate the number of sources \cite{ye2024}. In the simulations, we use E-AML with Alternate Projection capped at a maximum of $100$ iterations, which we empirically found that does not worsen the RMSE. Since the authors do not provide any codebook, we simulated E-AML with the proposed codebook in this paper. For benchmarking purposes, in the first set of simulations we consider the same setup as in \cite{matiwax1999}, considering an ULA with $\nx=8$ sensors and $\nrf\in\{2,4\}$ RF-chains. The number of snapshots is fixed at $K=192$. Thus, the codebook size yields  $M=2\nx/\nrf\in\{8,4\}$ (BSA and PI), and   $M=\ceil{\nx/(\nrf-1)}\in\{8,3\}$ (LS 2DFT, Cl-GLS and E-AML). 
The number of snapshots is fixed at $K=192$. Thus, the codebook size yields  $M=2\nx/\nrf\in\{8,4\}$ (BSA and PI), and   $M=\ceil{\nx/(\nrf-1)}\in\{8,3\}$ (LS 2DFT, Cl-GLS and E-AML)


\subsection{ULA: RMSE vs DoA}
\begin{figure}[ht]
    \centering
    \subfloat[][\label{fig:rmse_vs_theta}]{%
        \includegraphics[width=0.485\textwidth]{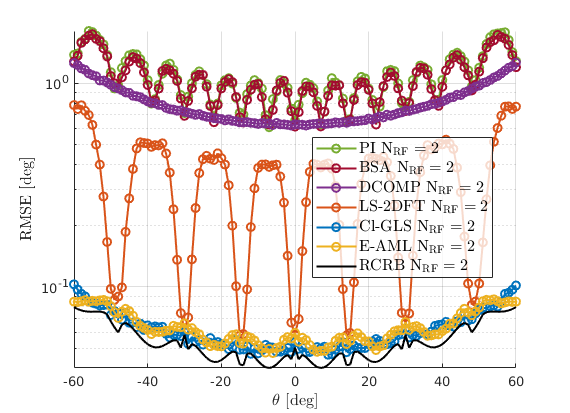}%
    }%

    \subfloat[][\label{fig:presolution_vs_delta}]{%
        \includegraphics[width=0.485\textwidth]{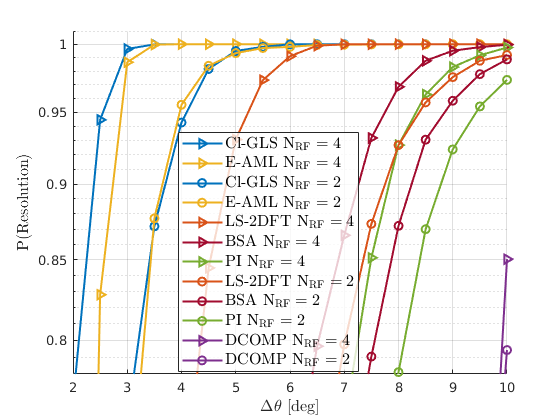}%
    }%

    \caption{(a) RMSE as a function of DoA. $N=8$, $\nrf=2$, $K=192$, $L=1$, $\text{SNR}=20 \;\text{dB}$. 
    (b) Probability of resolution as function of angular separation. $\nx=8$, $\nrf\in\{2,4\}$, $(\theta_0,\theta_1)=(0^\circ,\Delta\theta^\circ)$, $K=192$, $\text{SNR}=10\ \text{dB}$.}
    \label{fig:combined_plots}
\end{figure}

The first simulation analyzes the impact of the DoA in the different algorithms, by considering a single source with $\text{SNR}=20 \text{ dB}$. As illustrated in Fig.~\ref{fig:rmse_vs_theta}, the proposed Cl-GLS outperforms all competing algorithms techniques and it has similar performance than E-AML. In addition to improved accuracy, it yields a smoother RMSE profile with significantly reduced variability. By contrast, methods such as BSA, PI, and 2D-DFT LS exhibit highly irregular RMSE curves characterized by pronounced peaks and valleys, with the valleys typically located at the centers of the DFT beams. This behavior aligns with expectations, as the GLS framework incorporates a statistical weighting mechanism that effectively whitens the error contributions across beams, thereby enhancing estimation consistency over the angular domain. Meanwhile, DCOMP presents an even more uniform RMSE profile, as the pseudo-random nature of its beamforming matrices inherently smooths out the RMSE.


\begin{figure}[ht]
    \centering
    \subfloat[]{\includegraphics[width=0.47\textwidth]{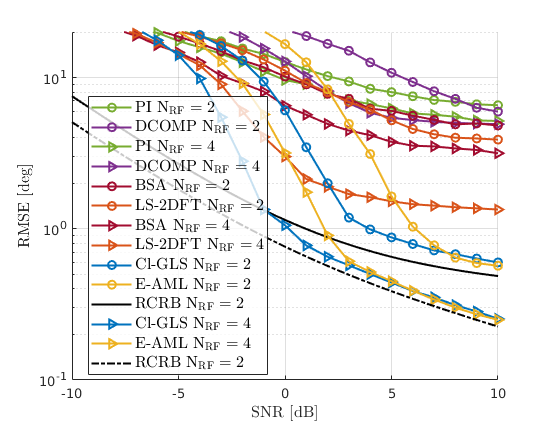}\label{fig:rmse_vs_snr}}
    
    \subfloat[]{\includegraphics[width=0.47\textwidth]{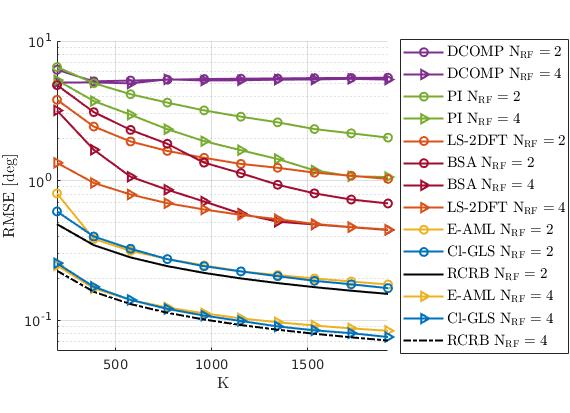}\label{fig:rmse_vs_K}}
        
    \caption{ (a) RMSE as a function of SNR. $\nx=8$, $\nrf\in\{2,4\}$, $(\theta_0,\theta_1)=(-2.56^\circ,2.56^\circ)$, $K=192$. 
    (b) RMSE as function of K. $\nx=8$, $\nrf\in\{2,4\}$, $(\theta_0,\theta_1)=(-2.56^\circ,2.56^\circ)$, $\text{SNR}=10\ \text{dB}$.}
\end{figure}
\subsection{ULA: PROBABILITY OF RESOLUTION vs ANGLE SEPARATION}
To evaluate the capability of the algorithm to separate closely spaced angular signals, we analyze the probability of resolution, $P(\text{Resolution})$\footnote{By definition, two signals are resolved if $|\hat{\theta}_0-\theta_0|$ and $|\hat{\theta}_1-\theta_1|$ are smaller than $|\theta_0-\theta_1|/2$ \cite{trees2002,stoica1999}.}, as a function of angular separation. This metric provides a measure of the spatial separation required for reliable estimation, as illustrated in Fig.~\ref{fig:presolution_vs_delta}. For these simulations, the SNR is fixed to $\text{SNR}=10\ \text{dB}$ with two signals impinging on the array at angles $(\theta_0,\theta_1)=(0^\circ,\Delta\theta^\circ)$. These results show that while the competing alternatives require an angular separation of more that $10^\circ$ to resolve the sources, both E-AML and the proposed Cl-GLS achieve perfect resolution $(P(\text{Resolution})=1)$ at approximately $6^\circ$. Although the overall performance of E-AML and Cl-GLS is comparable, Cl-GLS converges to perfect resolution at slightly smaller angular separation.

\subsection{ULA: RMSE vs SNR}
To evaluate the impact of the SNR on the performance of the different methods, we also adopt the setup of \cite{matiwax1999}, where two closely spaced signals impinge on the array from DoAs $(\theta_0,\theta_1)=(-2.56^\circ,2.56^\circ)$. Fig.~\ref{fig:rmse_vs_snr} shows that Cl-GLS outperforms other approaches based on covariance estimation. Of particular importance is the dependence of the RMSE on the SNR. It is well documented that when the number of RF chains is lower than the number of sources ($\nrf \leq L$), the RMSE stops decreasing for high SNR  \cite{sheinvald1997,matiwax1999}, as we can note for the case $\nrf=2$. E-AML displays minimal improved performance in high SNR with respect to Cl-GLS, approximately $0.03^\circ$ for $\nrf=2$ and $0.01^\circ$ for $\nrf=4$. Conversely, the remaining baseline methods yield poor performance across the entire SNR range. As demonstrated Fig.~\ref{fig:presolution_vs_delta}, this stems from their inability to resolve closely spaced sources, making them ineffective in this scenario where the angular separation is only $5.12^\circ$.


\subsection{ULA: RMSE vs SNAPSHOTS}
In the fourth simulation, we analyze the performance of the algorithm as the number of snapshots increases. The setup is configured with $N = 8$, $\nrf = {2, 4}$, and DoAs $(\theta_0, \theta_1) = (-2.56^\circ, 2.56^\circ)$, with $\text{SNR} = 10 \ \text{dB}$. The results are illustrated in Fig. \ref{fig:rmse_vs_K}. Similarly to previous examples, our method outperforms state-of-the-art algorithms. Unlike the earlier experiments, the other methods do not exhibit an error floor as the number of snapshots increases. In fact, for the 2D-DFT LS method, the RMSE continues to decrease with more snapshots. However, compared to these methods, Cl-GLS consistently delivers significantly better results. Notably, as the number of snapshots increases, the RMSE of our method approaches the CRB. This behavior is attributed to the improved accuracy of the estimators ${\hat{\bm S}_m}$, whose errors diminish with additional samples.



\begin{figure}[t!]
    \centering
    \subfloat[][\label{fig:rmse_vs_L}]{%
        \includegraphics[width=0.485\textwidth]{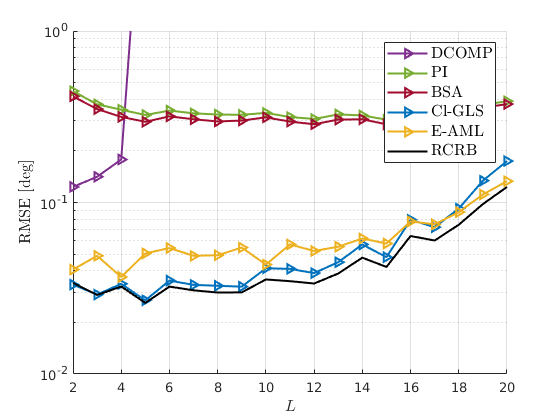}%
    }%

    \subfloat[][\label{fig:time_vs_L}]{%
        \includegraphics[width=0.485\textwidth]{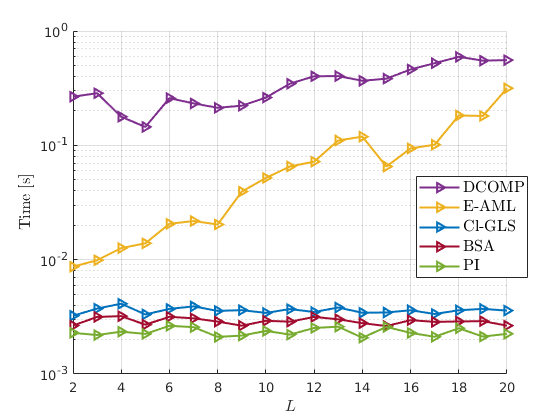}%
    }
    
    \caption{Array parameters: $N=32$, $\nrf=4$, $\text{SNR}=10\ \text{dB}$ and $K=352$. $L$ is the number of sources.  (a) RMSE vs $L$. (b) Time vs $L$.}
\end{figure}

\begin{figure}[t!]
    \centering
    \subfloat[][\label{fig:ura_rmse_vs_snr_theta}]{%
        \includegraphics[width=0.485\textwidth]{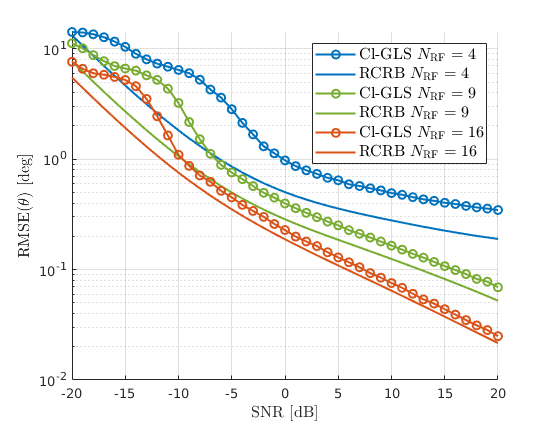}%
    }%

    \subfloat[][\label{fig:ura_rmse_vs_snr_phi}]{%
        \includegraphics[width=0.485\textwidth]{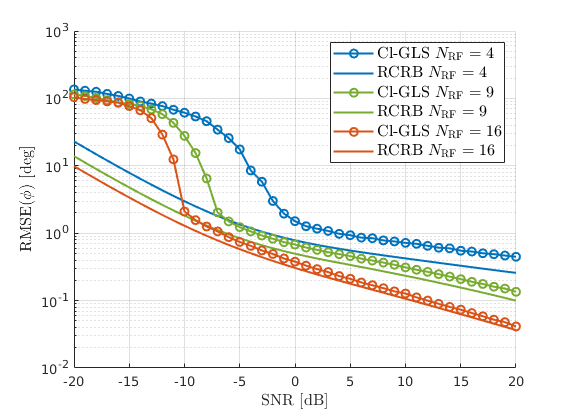}%
    }
    
    \caption{Comparison of $\text{RMSE}$ vs $\text{SNR}$.
    Simulation parameters: $\nx=\ny=6$, $\nrf\in\{4,9,16\}$, $L=4$,  $K=720$. (a) RMSE for elevation; (b) RMSE for azimuth.}
    \label{fig:ura_rmse_vs_snr}
\end{figure}
\subsection{ULA: RMSE vs NUMBER OF SOURCES}
For this benchmarking, we consider a ULA with $N=32$ antennas, $\nrf=4$ RF-chains, $\text{SNR}=10\ \text{dB}$ and $K=352$ snapshots. We vary the number of sources $L\in\{2,\dots,20\}$, and for each configuration, the DoAs are chosen to be uniformly spaced within $[-60^{\circ},60^{\circ}]$ according to $\theta_{\ell} = -60^{\circ}+\ell\cdot 120^{\circ}/(L-1)$ for $0\leq\ell<L$. Note that the irregular pattern observed in the RMSE curves in Fig.~\ref{fig:rmse_vs_L} is due to the fact that the RMSE depends on the specific DoAs as previously illustrated in Fig.~\ref{fig:rmse_vs_theta}. Furthermore, it is worth noting that a direct implementation of the alternating projection (AP) algorithm for E-AML converges to incorrect local minima introducing severe outliers that skew the overall RMSE measurements as noted in \cite{barthelme2021_trans}. To mitigate this problem, we discard these anomalies by evaluating the RMSE using only the top $90\%$ of the best realizations of this algorithm.
We observe that Cl-GLS consistently outperforms all the other methods up  to $L=17$, where E-AML begins to dominate. However, the latter scheme entails a high computational cost.  To quantify this, we used the MATLAB\textsuperscript{\textregistered} \textsc{timeit} function to measure the execution time of both methods, with important savings for Cl-GLS, more remarkable for higher $L$. As demonstrated in Fig.~\ref{fig:time_vs_L}, and for $L=20$, E-AML takes 75 more time to execute than Cl-GLS.

\subsection{URA: RMSE vs SNR}
For the simulations of the uniform rectangular array, we consider a receiver with a URA consisting of $\nx=\ny=6$ antennas, with inter-element spacing $d_x=d_y=\frac{\lambda}{2}$. The number of RF-chains is the same for both horizontal and vertical dimensions: $\nrfx=\nrfy=\sqrt{\nrf}$, with $\nrf \in \{4, 9, 16\}$. After estimating the SCM, we use Unitary-ESPRIT algorithm \cite{trees2002} to estimate the DoAs. The performance metric is the RMSE as defined in \eqref{eq:RMSE} for elevation $(\text{RMSE}(\theta))$ and azimuth $(\text{RMSE}(\phi))$.

We evaluate the impact of the SNR on the performance of the algorithm.  We assume $L=4$ uncorrelated signals impinging on the array from the directions $(\theta,\phi) = \{(30^{\circ}, 30^{\circ}), (35^{\circ}, 40^{\circ}), (45^{\circ}, 80^{\circ}), (55^{\circ}, 160^{\circ})\}$. The number of temporal snapshots is fixed at $K = 720$, while the number of snapshots per batch, $\Km$, varies depending on the number of RF chains. Elevation and azimuth results are  presented in Figs. \ref{fig:ura_rmse_vs_snr_theta} and \ref{fig:ura_rmse_vs_snr_phi}, respectively.  In both cases, the proposed method demonstrates performance close to the CRB, particularly as the number of RF chains increases. Notably, even with a limited number of RF chains, the method maintains good performance, indicating its applicability in hardware-constrained scenarios.

\section{CONCLUSIONS}
In this work, we have introduced a novel framework for covariance matrix reconstruction in Hybrid Analog and Digital array architectures with direct application to Direction-of-Arrival estimation. By exploiting the Cauchy-like displacement structure inherent in the signal after DFT beamforming, we established a rigorous linear relationship between the compressed measurements and the Sample Spatial
Covariance Matrix of the equivalent Fully-Digital array.

The proposed approach is centered on a Generalized Least Squares estimator that exploits the information present in the cross-correlations between the different RF-chains outputs. By including these 
measurements, the estimator achieves near efficient performance. To reduce the computational complexity, we leveraged the specific algebraic structure of the linear relationship to derive a computationally efficient implementation with Uniform Linear Arrays, reducing the complexity from $\mathcal{O}(\nrf^2\nx^3)$ to $\mathcal{O}(\nrf^2\nx)$.

Finally, we demonstrate that this methodology provides a seamless extension to Uniform Rectangular Arrays (URAs). While the URA framework established a theoretical baseline, the fast computational technique developed for the linear array cannot be directly mapped to URAs, identifying a promising line for future research.


\begin{appendices}
\makeatletter
\renewcommand{\thesubsection}{\thesection.\arabic{subsection}}
\renewcommand{\thesubsectiondis}{\thesection.\arabic{subsection}.}
\makeatother

\section{MATRIX DEFINITIONS}\label{appendix:A}
\subsection{CAUCHY-LIKE OPERATOR \texorpdfstring{$\Sigmacl$}{}}\label{appendix:A1}
This Appendix provides the definition of the constituent matrices $\{\bm{D}_n\}$ of the Cauchy-like operator 
\begin{equation}
    \Sigmacl=[\vec(\bm D_0),\dots,\vec(\bm D_{2\nrf-2})],
\end{equation}
introduced in \eqref{eq:Sigmacl}, and proves its full column rank. 
The basis matrices $\{\bm{D}_n\}_{n=0}^{2\nrf-2}$ are defined   as  
\begin{equation}\label{eq:Deven}
    \bm{D}_{2l} \triangleq \alpha_{l,l} \bm{E}_{l,l}\quad 0\leq l<\nrf,
\end{equation}
\begin{equation}
    \bm{D}_{2l+1} \triangleq \sum\nolimits_{u=0}^{l}\sum\nolimits_{v=l+1}^{\nrf-1}\left(\alpha_{u,v}\bm{E}_{u,v}-\alpha_{v,u}\bm{E}_{v,u}\right),
\end{equation}
for even and odd indexes, respectively, with $0\leq l<\nrf-1$. The weights $\alphauv$ introduced in \eqref{eq:alpha_uv} satisfy
\begin{equation}\label{eq:Dodd}
    \alphauv = \begin{cases}
        2 & u=v\\
        \frac{2j}{\nx}(1-e^{j\frac{2\pi}{\nx}(v-u)})^{-1} & u\neq v.
    \end{cases}
\end{equation}

\begin{lemma}
    The matrix $\Sigmacl\in\mathbb{C}^{\nrf^2\times(2\nrf-1)}$ has full column rank, i.e., $\rank(\Sigmacl)=2\nrf-1$.
\end{lemma}
\begin{proof}
To establish the linear independence of the columns of $\Sigmacl$, it suffices to show that a linear combination of its columns vanishes if and only if all scalar weights are identically zero. Let $\{\beta_n\}_{n=0}^{2\nrf-2}$ be a set of scalars. By separating the summation into even- and odd-idexed terms, the condition for this linear combination to equal zero yields:
 \begin{align}\label{eq:Sigmacl_linear_independence}
    \sum_{l=0}^{\nrf-1}\beta_{2l}\bm D_{2l} + \sum_{l=0}^{\nrf-2}\bm \beta_{2l+1}\bm D_{2l+1}=\bm 0_{\nrf\times \nrf}.
\end{align}
Observe that the even-index matrices $\{\bm D_{2l}\}$ contains non-zero entries only along the main diagonal, whereas the odd-index matrices $\{\bm D_{2l+1}\}$ contain non-zero entries exclusively in the off-diagonal positions.
We first analyze the even-index matrices. From \eqref{eq:Deven}, $\bm D_{2l}=2\cdot\bm E_{l,l}$, and therefore the diagonal entries in \eqref{eq:Sigmacl_linear_independence} vanish only if $\beta_{2l}=0$ for $0\leq l<\nrf$. Next, consider the odd-index matrices $\{\bm D_{2l+1}\}$. By construction, the entry located at position $(l,l+1)$ is non-zero only for the matrix $\bm D_{2l+1}$. More precisely, $\bm D_{2l'+1}[l,l+1]=0$ when $l'\neq l$, whereas $\bm D_{2l+1}[l,l+1]=\alpha_{l,l+1}\neq 0$. Hence, the off-diagonal entries in \eqref{eq:Sigmacl_linear_independence} vanish only if $\beta_{2l+1}=0$ for $0\leq l<\nrf-1$. Since satisfying \eqref{eq:Sigmacl_linear_independence} requires every scalar $\beta_n$ to be zero, the columns of $\Sigmacl$ are linearly independent, confirming that $\rank(\Sigmacl)=2\nrf-1$.\end{proof}

\subsection{SPECTRAL OPERATOR \texorpdfstring{$\bm \Omega$}{}}\label{appendix:A2}
The entries of the square matrix $\bm \Omega\in\mathbb{C}^{(2\nx-1)\times(2\nx-1)}$ read as
\begin{align}\label{eq:Omega}
\mathbf{\Omega}&[u,q+\nx-1] \triangleq \begin{cases}
\frac{1}{2}\left(1 - \frac{|q|}{N_x}\right)e^{-j\frac{\pi}{N_x} uq}, & u \text{ is even}, \\
\sin\left(\frac{\pi|q|}{N_x}\right)e^{-j\frac{\pi}{N_x} uq}, & u \text{ is odd},
\end{cases}\nonumber \\
& 0 \leq u < 2\nx-1, \quad -\nx+1 \leq q < \nx.
\end{align}
\begin{lemma}
    The matrix $\bm \Omega\in\mathbb{C}^{(2\nx-1)\times(2\nx-1)}$ is full rank, i.e., $\rank(\bm\Omega)=2\nx-1.$
\end{lemma}
\begin{proof}
    To prove that $\bm \Omega$ is full rank, we must demonstrate that its columns are linearly independent. That is, for a set of scalar coefficients $\{\beta_q\}_{q=-\nx+1}^{\nx-1}$, the linear combination
    \begin{equation}\label{eq:Omega_linear_independence}
    \sum\nolimits_{q=-\nx+1}^{\nx-1}\beta_q\cdot \bm \Omega[u,q+\nx-1]=0\quad \forall u,
\end{equation}
holds if and only if $\beta_q=0$ for all $q$.
Since $\bm \Omega$ is defined piecewise for even and odd rows, we decouple the system by evaluating the row indices $u=2u'$ and $u=2u'+1$. Moreover, within the summation range $-\nx+1\leq q<\nx$, the complex exponential exhibits periodicity. Specifically, for the index $0\leq q'<\nx$, they satisfy $e^{-j\frac{2\pi}{\nx}u'q'}=e^{-j\frac{2\pi}{\nx}u'(q'-\nx)}$. By aggregating the terms that share the same frequency, we rewrite \eqref{eq:Omega_linear_independence} for even and odd rows as
\begin{align}
    &\sum_{q'=0}^{\nx-1}\left[\tfrac{\beta_{q'}}{2}\Big(\tfrac{\nx-q'}{\nx}\Big)+\tfrac{\beta_{q'-\nx}}{2}\Big(\tfrac{\nx+q'}{\nx}\Big)\right]e^{-j\tfrac{2\pi}{\nx}u'q'}=0,\nonumber\\
    &\sum_{q'=1}^{\nx-1}\left[(\beta_{q'}-\beta_{q'-\nx})\sin\left(\tfrac{\pi q'}{\nx}\right)e^{-j\tfrac{\pi}{\nx}q'}\right]e^{-j\tfrac{2\pi}{\nx}u'q'}=0,\nonumber
\end{align}
where the odd sum starts at $q'=1$ as $\sin(0)=0$. For these summations to be zero, the terms inside the square brackets must vanish, as they represent the coefficient of a linearly independent Fourier basis. Equating the bracketed expressions to zero simplifies into the equations: $(\nx-q')\beta_{q'}+q'\beta_{q'-\nx}=0$ and $\beta_{q'}-\beta_{q'-\nx}=0$.
Evaluating the system at $q'=0$ yields $\nx\beta_0=0$, which implies $\beta_0=0$. For the remaining indices $1\leq q'<\nx$, substituting the odd-row condition $(\beta_{q'}=\beta_{q'-\nx})$ into the even-row equation dictates that $\beta_{q'}=\beta_{q'-\nx}=0$. Consequently, $\beta_q=0$ for all $q$. This confirms that the columns are linearly independent, and thus the matrix $\bm \Omega$ is full rank.
\end{proof}
\section{STAGES FOR GLS FAST IMPLEMENTATION}\label{appendix:B}

This Appendix provides the comprehensive mathematical derivations supporting the fast GLS implementation introduced in Section~\ref{sec:gls_fast_implementation} and summarized in Algorithm~\ref{alg:pseudocode_fast_solver}. Specifically,  Appendices~\ref{appendix:B1}--\ref{appendix:B4} correspond to the operations required for each of the four stages. Recall that our objective is to develop a computational efficient implementation to evaluate the estimate
\begin{equation}
    \hatrx = \bm \Omega^{-1}(\bm \Psi^H\hatReps^{-1}\bm \Psi)^{-1}\bm \Psi^H\hatReps^{-1}\hatp,
\end{equation}
 where $\bm \Psi=(\bm I_M\otimes\Sigmacl)\Pf\bm\Phi$. We also note that the matrices $\Pf$ \eqref{eq:Pf} and $\bm \Phi$ \eqref{eq:Phi} are highly sparse; consequently, matrix products involving these operators can be computed with minimal overhead.

In summary, Appendix~\ref{appendix:B1} details the steps to evaluate $(\bm I_M\otimes\Sigmacl^H)\hatReps^{-1}(\bm I_M\otimes\Sigmacl^H)$ and $(\bm I_M\otimes\Sigmacl^H)\hatReps^{-1}\hatp$; Appendix~\ref{appendix:B2} demonstrates that the product $\bm \Psi^H\hatReps^{-1}\bm \Psi$ can be expressed as a banded matrix plus a low-rank update $\bm\Psi^H\hatReps^{-1}\bm\Psi=\bm Q+\bm U\bm C\bm U^H$; Appendix~\ref{appendix:B3} covers the estimation of the spectral vector $\hatscl=(\bm \Psi^H\hatReps^{-1}\bm\Psi)^{-1}\bm\Psi^H\hatReps^{-1}\hatp$ by applying the Woodbury Inversion Lemma; and Appendix~\ref{appendix:B4} describes the final transformation from $\hatscl$ to $\hatrx$ via the IFFT.
\subsection{STAGE I: BATCH CAUCHY-LIKE PROCESSING}\label{appendix:B1}
Since $\hatReps$ and $(\bm I_{M}\otimes\Sigmacl)$ are block diagonal matrices, their products preserve this partitioned structure. Consequently, the evaluation of the terms
\begin{align}\label{eq:block_mat_vec}
    &(\bm I_{M}\otimes \Sigmacl)^H\hatReps^{-1}\hatp = \\
    &=[(\Sigmacl^H\hat{\bm R}_{\epsilon,0}^{-1}\vec(\hat{\bm{S}}_0))^T, \dots,(\Sigmacl^H\hat{\bm R}_{\epsilon,M-1}^{-1}\vec(\hat{\bm{S}}_{M-1}))^T]^T,\nonumber
\end{align}
and 
\begin{align}\label{eq:block_mat_mat}
    &(\bm I_{M}\otimes \Sigmacl)^H\hatReps^{-1}(\bm I_{M}\otimes \Sigmacl)^H = \nonumber\\
    &=\text{blk diag}(\Sigmacl^H\hat{\bm R}_{\epsilon,0}^{-1}\Sigmacl,\dots,\Sigmacl^H\hat{\bm R}_{\epsilon,M-1}^{-1}\Sigmacl),
\end{align}
decouples into a set of $M$ independent products, where the dimensions are dictated by the number of RF-chains ($\nrf$).
In typical hybrid architectures, the number of RF-chains is  significantly smaller than the number of antennas. However, because the Kronecker product $\hatRepsm=\Km^{-1}(\hatSm^{-T}\otimes\hatSm^{-1})$ yields a matrix of dimensions $\nrf^2\times\nrf^2$, a naive evaluation of the matrix-vector product $\Sigmacl^H\hatRepsm^{-1}\vec(\hatSm)$ and the matrix product $\Sigmacl^H\hatRepsm^{-1}\Sigmacl$ would incur complexity of $\mathcal{O}(\nrf^4)$ and $\mathcal{O}(\nrf^5)$, respectively. Hence, even for systems with a moderate number of RF-chains, this computational burden often becomes prohibitively high. To circumvent this, we leverage the algebraic properties of the Kronecker product to bypass the explicit formation of $\hatRepsm$ entirely. Specifically, by invoking the Kronecker identity $\vec(\bm{A}\bm{B}\bm{C}) = (\bm{C}^T\otimes\bm{A})\vec(\bm{B})$, for matrices $\bm A$, $\bm B$ and $\bm C$, the expression $\Sigmacl^H\bm{R}_{\epsilon,m}^{-1}\vec(\hat{\bm{S}}_m)=\Km^{-1}\Sigmacl^H(\Sm^{-T}\otimes \Sm^{-1})\vec(\hat{\bm{S}}_m)$ admits the reformulation:
\begin{align}    
\Sigmacl^H\bm{R}_{\epsilon,m}^{-1}\vec(\hat{\bm{S}}_m) =\Km^{-1}\Sigmacl^H\vec(\hat{\bm{S}}_m^{-1}).
\end{align}
Recall from \eqref{eq:Sigmacl} that the columns of $\Sigmacl$ consist of a set of $2\nrf-1$ structured vectors $\{\vec(\bm{D}_n)\}$ (see Appendix~\ref{appendix:A}). By applying the identity $\vec(\bm{A})^H\vec(\bm B) = \text{trace}(\bm{A}^H\bm{B})$, we can express the products as Frobenius inner products:
\begin{align}
    [\Sigmacl^H\vec(\hatSm^{-1})]_{u} &= \text{trace}(\bm{D}_u^H\hatSm^{-1}),\label{eq:trace0}\\
    [\Sigmacl^H(\hatSm^{-T}\otimes\hatSm^{-1})\Sigmacl]_{u,v} &= \text{trace}(\bm{D}_u^H\hatSm^{-1}\bm{D}_v\hatSm^{-1}),\label{eq:trace1}
\end{align}
for $0\leq u,v<2\nrf$. Both traces in \eqref{eq:trace0} and \eqref{eq:trace1} can be evaluated with high efficiency by exploiting the displacement structure in $\{\bm{D}_n\}$, thus  reducing the complexity of the batch-level products $\Sigmacl^H\hatRepsm^{-1}\vec(\hatSm)$ and $\Sigmacl^H\hatRepsm^{-1}\Sigmacl$ from $\mathcal{O}(\nrf^4)$ and $\mathcal{O}(\nrf^5)$ to $\mathcal{O}(\nrf^2)$ and $\mathcal{O}(\nrf^3)$, respectively. 

Specifically, the difference $\bm D_{2l+1}-\bm{D}_{2(l-1)+1}$ for $1\leq l <\nrf-1$ admits the following rank-2 representation:
\begin{align}\label{eq:displacement_struct}
    &\bm{D}_{2l+1} - \bm{D}_{2(l-1)+1} =\sum\nolimits_{u=0}^{l-1}\left(\alpha_{u,l}\bm{E}_{u,l}-\alpha_{l,u}\bm{E}_{l,u}\right) \\
    &+ \sum\nolimits_{v=l+1}^{\nrf-1}\left(\alpha_{l,v}\bm{E}_{l,v}-\alpha_{v,l}\bm{E}_{v,l}\right)= \bm{e}_l^{\nrf}\alphal^H + \alphal \bm{e}^{\nrf,T}_l\nonumber
\end{align}
where the vector $\alphal\in\mathbb{C}^{\nrf\times 1}$ is defined entry-wise as
\begin{equation}
    \alphal[u] \triangleq \begin{cases}
        -\alpha_{u,l} & u\neq l,\\
        0 & u=0,
    \end{cases}
\end{equation}
for $0\leq u <\nrf$. Using this recursive relation, we can evaluate the required traces with complexity  $\mathcal{O}(\nrf^2)$ and $\mathcal{O}(\nrf^3)$, respectively.
\subsection*{\textit{Evaluation of $\tr{\bm{D}_u^H\hatSm^{-1}}$}}
Using the trace identity $\tr{\bm E_{u,v}\hatSm^{-1}}=\hatSm^{-1}[v,u]$, the even-indexed terms simplify to
\begin{equation}
    \text{tr}[\bm D_{2l}^H\hatSm^{-1}] = 2\hatSm^{-1}[l,l] \quad l=0,\dots,\nrf-1.
\end{equation}
The odd-indexed components $\{\tr{\bm D_{2l+1}^H\hatSm^{-1}}\}$ are computed recursively. We initialize the recursion by evaluating
\begin{equation}
    \tr{\bm D_{1}^H\hatSm^{-1}} = 2\text{Re}[\hatSm^{-1}[0,:]\bm{\alpha}_0].
\end{equation}
Applying the displacement relation in \eqref{eq:displacement_struct}, the subsequent terms are found via
\begin{equation}
    \tr{\bm D_{2l+1}^H\hatSm^{-1}} = \tr{\bm D_{2(l-1)+1}^H\hatSm^{-1}}- \alphal^H\hatSm^{-1}[:,l]+\hatSm^{-1}\alphal.\nonumber
\end{equation}
Computing all traces requires $\mathcal{O}(\nrf^2)$ operations, as each recursive step involves only vector-level multiplication and addition.
\subsection*{\textit{Evaluation of $\tr{\bm{D}_u^H(\hatSm^{-T}\otimes\hatSm^{-1})\bm D_{v}}$}}
To evaluate the second-order products, we define the auxiliary matrices  $\bm W_u \triangleq \hatSm^{-1}\bm D_u \hatSm^{-1}$, such that the trace becomes $\tr{\bm D_u^H\bm{W}_v}$. Using the properties $\hatSm^{-1}\bm E_{u,v}\hatSm^{-1}=\hatSm^{-1}[:,u]\hatSm^{-1}[v,:]$, the even-indexed matrices are given by
\begin{equation}
    \bm{W}_{2l} = \alpha_{l,l} \hatSm^{-1}[:,u]\hatSm^{-1}[u,:].
\end{equation}
The odd-indexed matrices $\{\bm W_{2l+1}\}$ are computed via recursive update. The initial component is
\begin{equation}
    \bm W_{1} = \sum_{v=0}^{\nrf-1}(\alpha_{0,v}\hatSm^{-1}[:,0]\hatSm^{-1}[v,:]-\alpha_{v,0}\hatSm^{-1}[:,v]\hatSm^{-1}[0,:]),\nonumber 
\end{equation}
and the remaining terms are updated as:
\begin{equation}
    \bm{W}_{2l+1} = \bm{W}_{2(l-1)+1} + \hatSm^{-1}[:,l](\alphal^H\hatSm^{-1}) + \hatSm^{-1}(\alphal\hatSm^{-1}[l,:]).\nonumber
\end{equation}
Since each update requires only two outer products, the calculation of the full set $\{\bm W_v\}$ (and consequently the required traces), attains a total complexity of $\mathcal{O}(\nrf^3)$.

\subsection{STAGE II: BANDED AND LOW-RANK DECOMPOSITION}\label{appendix:B2}
In this stage, the block-diagonal matrix $(\bm I_M\otimes \Sigmacl)^H\hatReps^{-1}(\bm I_M\otimes \Sigmacl)$ is already available from Stage I. While direct multiplication by operators $\Pf$ and $\bm \Phi$ yields a dense matrix that is computationally expensive to invert, we leverage the specific block structures of these operators to decompose the product into a banded matrix and a low-rank update. Specifically, the banded component $\bm Q$ has a bandwidth of $2\nrf-1$ produced by the staircase-like structure of $\Pf$ (see Fig.~\ref{fig:Pf}). The low-rank update, $\bm U\bm C\bm U^H$, accounts for the dimensionality expansion introduced by $\bm \Phi$ and the wrap-around effect of the final rows of $\Pf$.

To formalize this decomposition, let $\rdft$ and $\rwa$ denote the number of rows of $\Pf\in\{0,1\}^{M(2\nrf-1)\times 2\nx}$ associated with the measurement of the $2\nx$-DFT frequency grid and wrap-around blocks as 
\begin{equation}
    \rdft\triangleq 2\nx + M-1,\quad  \rwa\triangleq2[M(\nrf-1)-\nx]+1.
\end{equation}
Then we define the partition of the expansion matrix and frequency selection matrix as
\begin{equation}
\bm \Phi=\begin{bmatrix}
        \bm{I}_{2\nx-1}\\
        \bm v^T
    \end{bmatrix}, \quad \Pf = \begin{bmatrix}
        \bm{P}_{00}&\bm{e}_{\rdft-1}^{\rdft} \\
        \bm{P}_{10}&\bm{0}_{\rwa\times \rwa}
    \end{bmatrix}.
\end{equation}
By partitioning the result of the Stage I accordingly with the blocks of $\Pf$ as
\begin{equation}\label{eq:M_ij}
    (\bm{I}_M\otimes\Sigmacl)^H\hatReps^{-1}(\bm{I}_M\otimes\Sigmacl) = \begin{bmatrix}
        \bm{M}_{00}&\bm{M}_{01}\\
        \bm{M}_{10}&\bm{M}_{11}
        \end{bmatrix},
\end{equation}
we can express the following product
\begin{align}
    \bm \Phi^T\Pf^T&(\bm I_{M}\otimes\Sigmacl)^H\hatReps^{-1}(\bm I_{M}\otimes\Sigmacl)\Pf\bm \Phi= \bm Q + \bm U\bm C\bm U^H,\nonumber
\end{align}
where the constituent matrices $\bm Q\in\mathbb{C}^{(2\nx-1)\times(2\nx-1)}$, $\bm U\in\mathbb{C}^{(2\nx-1)\times 2(\rwa+1)}$ and $\bm C\in\mathbb{C}^{2(\rwa-1)\times 2(\rwa+1)}$ read:
{ 
\fontsize{9.5pt}{11pt}\selectfont 
\addtolength{\jot}{8pt} 
\begin{align}
    \bm{Q} &= \bm{P}_{00}^T\bm{M}_{00}\bm{P}_{00}, \label{eq:Q} \\
    \bm{U} &= [\bm{v},(\bm{P}_{00}^T\bm{M}_{00} + \bm{P}_{10}^T\bm{M}_{10})\bm{e}_{\rdft}^{\rdft-1},\bm{P}_{10}^T, \bm{P}_{00}^T\bm{M}_{01}], \label{eq:U} \\
    \bm{C} &= \blkdiag \left(
    \begin{bmatrix}
        \bm{M}_{00}[\rdft,\rdft] & 1 \\
        1 & 0
    \end{bmatrix}, 
    \begin{bmatrix}
        \bm{M}_{11} & \bm{I}_{\rwa} \\
        \bm{I}_{\rwa} & \bm{0}_{\rwa,\rwa}
    \end{bmatrix}
    \right).\label{eq:C}
\end{align}
}
Due to  the staircase-like structure of $\bm{P}_{00}$ and the block-diagonal form of $\bm{M}_{00}$, the product $\bm{Q}=\bm{P}_{00}^T\bm{M}_{00}\bm{P}_{00}$ is Hermitian banded, with a bandwidth of $2\nrf-1$.

\textit{Remark:} If $\nrf = \nx$, then the product $\Pf\bm{\Phi}$ reduces to the identity matrix $\bm{I}_{2\nx-1}$. In that situation, Stage II does not provide any computational advantage.
\subsection{STAGE III: ESTIMATION OF \texorpdfstring{$\hatscl$}{} VIA WOODBURY INVERSION LEMMA}\label{appendix:B3}
From Stage I and II, the components $(\bm I_M\otimes\Sigmacl^H)\hatReps^{-1}\hatp$, $\bm Q$ and $\bm U\bm C\bm U^H$ are available. We now address the efficient evaluation of the spectral vector estimate
\begin{equation}\label{eq:fast_solver_second_stage}
\hatscl = (\bm \Psi^H\hatReps^{-1}\bm \Psi)^{-1} \bm\Psi^{H}\hatReps^{-1}\hat{\bm{p}},
\end{equation}
where $\bm \Psi=(\bm I_{M}\otimes\Sigmacl)\Pf\bm\Phi$. First, the matrix-vector product $\bm \Phi^T\Pf^T(\bm I_M\otimes\Sigmacl^H)\hatReps^{-1}\hatp$ is computed. This operation entails negligible computational cost due to the extreme sparsity of the selection and expansion operators $\Pf$ and $\bm \Phi$. Consequently, the primary computational bottleneck lies in the matrix inversion.

To circumvent a direct inversion, we exploit the decomposition derived in Stage II, which expresses the inner matrix products as a low-rank update of a Hermitian banded matrix: $\bm Q+\bm U\bm C\bm U^H$. The rank of the perturbation satisfies $\text{rank}(\bm U \bm C \bm U^H)\leq 2(\rwa+1)$, where the number of rows associated with the wrap-around $\rwa$ is typically very small. Consequently, the evaluation of $\hatscl$ can be extremely simplified by applying the Woodbury Inversion Lemma, which expresses the required inverse as $(\bm Q+\bm U\bm C\bm U^H)^{-1} 
    =\bm{Q}^{-1} - \bm{Q}^{-1} \bm{U} \left( \bm{C}^{-1} + \bm{U}^H \bm{Q}^{-1} \bm{U} \right)^{-1} \bm{U}^H \bm{Q}^{-1}$. In this formulation, the primary computational burden involves computing the matrix-matrix product $\bm Q^{-1}\bm U$ and the matrix-vector product $\bm Q^{-1}\bm \Phi^T\Pf^T(\bm I_M\otimes\Sigmacl^H)\hatReps^{-1}\hatp$. Since $\bm Q$ is a banded matrix, specialized banded solvers, such as those based on LU or Cholesky decomposition, can compute these terms with a complexity $\mathcal{O}(\nrf^2\nx)$ \cite{golub2013matrix}. The explicit steps to apply the Woodbury Inversion Lemma are shown in Algorithm~\ref{alg:pseudocode_fast_solver}.

\textit{Remark:} Because $\bm U$ is also highly sparse, the evaluation of the product $\bm Q^{-1}\bm U$ can be exceptionally efficient. While further complexity reductions could be achieved by exploiting this structure, such optimizations are out of the scope of this paper.

\subsection{STAGE IV: INVERSE FAST FOURIER TRANSFORM}\label{appendix:B4}
The final stage of the proposed algorithm involves the spectral transformation of $\hatscl$ to the spatial estimate $\hatrx$. While a direct evaluation of the matrix-vector product $\bm{\Omega}^{-1}\hatscl$ requires a complexity of $\mathcal{O}(\nx^2)$, by leveraging the windowed DFT structure of $\bm{\Omega}$, this can be reduced to $\mathcal{O}(\nx\log\nx)$ via the FFT.

Specifically, note the relationship between $\hatscl$ and $\hatrx$ can be established entry-wise as
\begin{align}\label{eq:hatscl_appendix}
    \hatscl[u]&=\sum_{q=-\nx+1}^{\nx-1}\begin{cases}
        \we[q]\hatrx[q]e^{-j\frac{\pi}{\nx}uq},\\
        \wo[q]\hatrx[q]e^{-j\frac{\pi}{\nx}uq},
    \end{cases}
\end{align}
where $\we[q]$ and $\wo[q]$ denote the even and odd windows, defined for $-\nx<q<\nx$ and zero elsewhere as
$\we[q] \triangleq 1/2(1-|q|/\nx)$ and $\wo[q] \triangleq \sin\left(\pi|q|/\nx\right)$. Observe that since the weight function $\we$ and $\wo$ are independent of the index $u$, we can efficiently obtain $\hatrx$ from $\hatscl$ via the inverse DFT. 
 Specifically, by denoting the IDFT $\hatscl$ as $\hat{\bm{X}} = \text{IDFT}[\hatscl]$, the elements of $\hat{\bm{X}}$ satisfy
\begin{align}
    &\hat{\bm{X}}[k] = \frac{1}{2\nx}\sum_{u=0}^{N-1}(\hatscl[2u]+e^{j\frac{\pi}{\nx}k}\hatscl[2u+1])e^{j\frac{2\pi}{\nx}uk}\nonumber\\
    &= \sum_{q=-\nx+1}^{\nx-1}\hat{\bm{r}}[k](\we[k]+e^{j\frac{\pi}{\nx}(k-q)}\wo[k])\sum_{u=0}^{\nx-1}\frac{e^{j\frac{2\pi}{\nx}u(k-q)}}{2\nx}\nonumber,
\end{align}
for $0\leq k < 2\nx$. Due to the orthogonality of the basis of complex exponentials, the inner sum collapses when $q-k=\{0,\pm \nx\}$. This leads to the following coupled linear relationship
\begin{equation}\label{eq:decoupled_systems}
    \begin{bmatrix}
        \hat{\bm{X}}[k]\\
        \hat{\bm{X}}[k+\nx]
    \end{bmatrix} = \bm{W}_k \begin{bmatrix}
        \hatrx[k]\\
        \hatrx[k-\nx]
    \end{bmatrix}\quad 0\leq k<\nx,
\end{equation}
where the $2\times 2$ coefficient matrices are:
\begin{equation}\label{eq:Wk}
    \bm{W}_k \triangleq \frac{1}{2}\begin{bmatrix}
        \we[k]+\wo[k]&\we[k-\nx]-\wo[k-\nx]\\
        \we[k]-\wo[k]&\we[k-\nx]+\wo[k-\nx]
    \end{bmatrix}.\nonumber
\end{equation}
To reconstruct the sequence $\hatrx$, we solve these $2\times 2$ systems. Given that the associated computational cost, $\mathcal{O}(\nx)$, can be neglected with respect to the complexity of the IFFT implementation of the IDFT,  $\mathcal{O}(\nx\log\nx)$, the  overall transformation is dominated by the efficiency of the IFFT.
\end{appendices}

\bibliography{ref}
\bibliographystyle{ieeetr}
\end{document}